\UseRawInputEncoding
\documentclass[twocolumn]{autart}   

\usepackage{amsmath}
\usepackage{algorithm}
\usepackage{algpseudocode}
\usepackage{amssymb}
\usepackage{color}
\usepackage{enumerate}
\usepackage{wrapfig}
\usepackage{graphicx}           

\usepackage[dvips]{epsfig}    
\usepackage[round, sort]{natbib}
\usepackage{hyperref}

\usepackage{booktabs}
\usepackage{multirow}
\usepackage{enumitem}

\hypersetup{
	colorlinks = true,
	citecolor = cyan,
}

\newtheorem{theorem}{Theorem}
\newtheorem{lemma}{Lemma} 
\newtheorem{remark}{Remark} 
\newtheorem{definition}{Definition} 
\newtheorem{assumption}{Assumption}

\newtheorem{problem}{Problem}

\allowdisplaybreaks
\begin{document}
\begin{frontmatter}

	\title{Distributed Data-driven Unknown-input Observers \\
		for State Estimation} 
	
	\thanks[footnoteinfo]{The work was supported in part by the National Key R\&D Program of China under Grant 2021YFB1714800, in part by the National Natural Science Foundation of China under Grants 61925303, U22B2058, 62088101, and in part of the Fundamental Research Funds for the Central Universities under Grant 2024CX06085.}
	\thanks[footnoteinfo]{This paper was not presented at any IFAC 
		meeting. }

	\author[Bit]{Yuzhou Wei},
    \author[Unipd]{Giorgia Disar{\`o}},
	\author[Bit]{Wenjie Liu}, 
	\author[Bit]{Jian Sun},   
    \author[Unipd]{Maria Elena Valcher},         
	\author[Bit]{Gang Wang}

	\address[Bit]{National Key Lab of Autonomous Intelligent Unmanned Systems, Beijing Institute of Technology, Beijing 100081, China}      
	\address[Unipd]{Dipartimento di Ingegneria dell’Informazione, Universit{\` a} di Padova, via Gradenigo 6B, 35131 Padova, Italy}             
	\address{Email:~\{weiyuzhou@bit.edu.cn,~giorgia.disaro@phd.unipd.it,~liuwenjie@bit.edu.cn,~sunjian@bit.edu.cn,\\ 
			meme@dei.unipd.it,~gangwang@bit.edu.cn\}}
		\maketitle

	\begin{abstract}
	Unknown inputs related to, e.g., sensor aging, modeling errors, or device bias, represent a major concern 
	in wireless sensor networks, as they degrade the state estimation performance. To improve the performance, unknown-input observers (UIOs) have been proposed. Most of the results available to design UIOs are  based on explicit system models, which can be difficult or impossible to obtain in real-world applications.  
	Data-driven techniques, on the other hand, have become a viable alternative for the design and analysis of unknown systems using only data. In this context, a novel  data-driven distributed unknown-input observer (D-DUIO) for unknown continuous-time linear time-invariant (LTI)  systems is developed, which requires solely some data collected offline, without any prior knowledge of the system matrices.  In the paper, first, a model-based approach to the design of a DUIO is presented.
	A sufficient condition for the existence of such a DUIO is recalled, and a new one is proposed, that is prone to a data-driven adaption.
	Moving to a data-driven approach, 
	it is shown that   under suitable assumptions on the input/output/state data collected from the continuous-time system, it is possible to   both claim the existence of a D-DUIO and to derive its   matrices   in terms of the  matrices of  pre-collected data.
	Finally, the efficacy of the  D-DUIO is illustrated by means of numerical examples. 
	\end{abstract}


\end{frontmatter}	
\section{Introduction}
\label{sec.intro}

In dynamical control systems, distributed state estimation (DSE) approaches play a vital role, and a multitude of well-established tools have been developed, including consensus Kalman-based filtering  \cite{olfati2005, chen2016weighted}, Luenberger-like consensus estimation \cite{millan2013sensor}, and distributed moving-horizon estimation \cite{BrouillonDRO}, to name a few. DSE has a wide range of real-world applications, including power system monitoring, cooperative tracking and localization, and smart transportation; see, e.g., \cite{ ahmad2017online,farina2010distributed} and references therein.

However, practical concerns about the deployment of DSE methods exist. For instance, unknown inputs caused by sensor aging, modeling errors, calibration bias, and/or external disturbances/attacks can lead to severe deterioration in estimation performance \cite{trimpe2014event,2018Comparing}. Among different tools to tackle the estimation problem  in the presence of unknown inputs, unknown-input observers (UIOs) have attracted recurring attention due to their geometric decoupling capabilities \cite{valcher1999state,nazari2019distributed}.
A distributed UIO (DUIO) was first implemented in \cite{chakrabarty2016distributed},  to estimate
the internal states of the nonlinear subsystems using local measurement
outputs.
More recently, a distributed UIO was developed for a continuous-time LTI system in \cite{yang2022state} by  resorting to a consensus strategy, in which the global system state is estimated consistently by each local observer with limited information about the input and output. 
A similar strategy was also employed in \cite{CaoWang}, but with a different structure for the  observer gain matrices, based on the  decomposition of the  state space  at each node into  detectability/undetectability subspaces.

It is worth noting that all of the previous results  about DUIOs
were derived
assuming that the  original
system models were known. However, obtaining accurate system models for interconnected cyber-physical systems, either  from first-principles or through system identification methods, is becoming increasingly difficult or even impossible. To address this challenge, data-driven control 
methods have gained attention in the big data era, aiming to design controllers directly from data without relying on intermediate system identification procedures, as described, e.g.,  in \cite{2013Model,depersis2020data}. Recent efforts leveraging \emph{Willems et al.}'s  fundamental lemma   \cite{willems2005note} have   addressed data-driven predictive control \cite{deepc,berberich2020data},
data-driven event-triggered and consensus control \cite{scis2023liyifei}, and data-driven observers \cite{Mishraddobserver}.
However, data-driven state estimation with unknown inputs has received only partial attention up to now. In \cite{Shireconstruction}, a data-driven input reconstruction method from outputs was developed to design inputs. The work  \cite{turan2021data}
investigated the data-driven UIO problem for unknown linear systems, 
with the goal of  estimating the state even in the presence of unknown disturbances. 
This work was recently extended in \cite{TAC-UIO}, where 
necessary and sufficient conditions for the problem solution, as well as a parametrization of all possible solutions, based only on data, were provided. 
Nevertheless, all these studies have only considered  centralized systems, and to the best of our knowledge, no results for data-driven DSE have been reported.

This paper aims to  fill this gap by developing a distributed data-driven  UIO scheme for  a continuous-time unknown linear system subject to unknown inputs and disturbances. Specifically, we introduce a novel data-driven DUIO (D-DUIO) designed using offline input/output/state data without performing any system identification. The D-DUIO with a consensus strategy allows the estimation of the unknown global system state through local information exchanges between neighboring nodes, even when no node has access to the complete input information. It is  shown that, under mild conditions, the local state estimates obtained by the nodes reach consensus and converge to the true state of the unknown system.

In summary, the contributions of this work are the following:
\begin{itemize}
	\item  By resorting to a model-based approach, we recall a sufficient condition - already derived in the literature - for the proposed DUIO to provide a state estimate that asymptotically converges to the true state of the system, and we propose a new sufficient condition, that, even if slightly stronger, is more suitable to be adapted to a data-driven context.
	\item By leveraging the results in \cite{turan2021data, TAC-UIO}, we provide necessary and sufficient conditions to verify using only the collected data whether the above mentioned sufficient condition for the existence of a DUIO is satisfied. 
	\item  We explicitly provide the data-driven expression of the matrices of the proposed D-DUIO. 
\end{itemize}

To make the paper flow smoother, all the proofs have been moved to the Appendix.

For convenience, we  introduce some notation. 
The sets of real numbers, nonnegative real numbers and nonnegative integers are denoted by $\mathbb{R}, {\mathbb R}_+,$ and $\mathbb{Z}_+$, respectively. 
$I_n$ denotes the identity matrix of size $n$, $\mathbf{0}_{m\times n}$ the zero matrix of size $m\times n$, and ${\mathbf 1}_n$ the all-one vector of size $n$. Suffixes will be omitted when the dimensions can be deduced from the context. 
The Moore-Penrose pseudoinverse of a matrix
$Q$  is denoted by $Q^{\dagger}$. 
We use $\ker(Q)$ to represent the kernel space of $Q$ and ${\rm range}(Q)$ to represent its column space. 
The spectrum of a square matrix $Q$ is denoted by $\sigma(Q)$ and is the set of all its eigenvalues. For a symmetric matrix $Q$, we use  $\lambda_{min}(Q)$ to denote the  smallest  eigenvalue of  $Q$. A symmetric matrix $Q$ is positive definite if $x^\top Q x >0$ for every $x\ne {\bf 0}$. When so, we adopt the notation $Q \succ 0$.

The Kronecker product is denoted by $\otimes$.
Given matrices $M_i, i\in \{1,2,\dots,p\}$,
the  block-diagonal matrix whose $i$th diagonal block is the matrix $M_i$ is denoted by ${\rm diag} (M_i)$. 
We also use  ${\rm diag} (M_i)_{i=2}^p$ to denote the block diagonal matrix with diagonal blocks $M_2, \dots, M_p$.

\section{Preliminaries and Problem Formulation}\label{sec.prosta}
The problem set-up we adopt is analogous to those  adopted  in  \cite{yang2022state} and \cite{CaoWang}. Specifically, we
consider a continuous-time LTI system
\begin{equation}\label{eq:system}
\dot x(t) = Ax(t) + Bu(t) + Ed(t), 
\end{equation}
where $t  \in {\mathbb R}_+$,
$x(t) \in \mathbb{R}^{n_x}$ is the state, $u(t) \in \mathbb{R}^{n_u}$ is the control input, $d(t) \in \mathbb{R}^{n_d}$ is the unknown process  disturbance,
$A \in \mathbb{R}^{n_x \times n_x}$, $B \in \mathbb{R}^{n_x \times n_u}$, and $E \in \mathbb{R}^{n_x \times n_d}$.

A wireless sensor network comprising $M$    heterogeneous  sensor nodes is deployed to monitor the state of system \eqref{eq:system}. At each time instant, each node of the network provides a  measured
output signal $y_i(t) \in \mathbb{R}^{n_{y_i}}$, given by
\begin{equation}	\label{eq:output}
	y_i(t) = C_ix(t),  \quad \forall i \in \mathcal{M} := \{1,2,\ldots,M\},
\end{equation}
where  $C_i \in \mathbb{R}^{n_{y_i} \times n_x}$.  
Moreover, 
we assume that each sensor node has access only to a subset of the input entries, and hence
for every $i \in \mathcal{M}$, we can split the entries of the control input $u(t)$ into two parts: the measurable part $u_i(t)$ and the unknown part $u_i^u(t)$. Consequently, we can always express $B u(t)$ as:
\begin{equation}	\label{eq:input}
	Bu(t) = B^m_i u_i(t) + B^u_i u_i^u(t),
\end{equation}
where $u_i(t) \in {\mathbb R}^{n_{m_i}}$, $B^m_i \in {\mathbb R}^{n_x\times n_{m_i}}$, $u_i^u(t) \in {\mathbb R}^{n_{p_i}}$, $B^u_i \in {\mathbb R}^{n_x\times n_{p_i}}$. Since $d(t)$ is also unknown for each node,  the overall unknown input at node $i$ and the associated system matrix can be represented as
\begin{subequations}	\label{eq:wandb}
	\begin{align}
		w_i(t) &:= \big[(u_i^u)^\top\!(t) \ \ d^\top(t)\big]^\top,\\
		B_i^p &:= [ B^u_i \ E ].
	\end{align}
\end{subequations}
Consequently, for every  $i \in \mathcal{M}$, 
the system dynamics, from the perspective of the $i$th sensor node, is given by:
\begin{equation}\label{eq:systemi}
	\dot x(t) = Ax(t) + B_i^m u_i(t) + B_i^p w_i(t).
\end{equation}
It is worthwhile noticing that the specific expression of $w_i(t)$ will play no role in the following, and hence we can always assume, possibly redefining $w_i(t)$, that the matrix $B_i^p$ is of full column rank $r_i := n_{p_i}+ n_d$ (see  \cite{yang2022state}). In the following, we will   denote by ${\mathcal T}_i$ the system described by the pair of equations 
\eqref{eq:system}--\eqref{eq:output} or, equivalently, by the pair \eqref{eq:systemi}--\eqref{eq:output}. 

The objective of each node is to reconstruct the global state $x(t)$ of the system, by exchanging information with other nodes. 
Specifically, we assume that
the sensor network is represented by a graph $\mathcal{G} = (\mathcal{M},\mathcal{E},\mathcal{A})$, where $\mathcal{M} =\{1,2,\dots, M\}$ is the set of sensor nodes, $\mathcal{E} \subseteq \mathcal{M}\times \mathcal{M}$ is the set of communication links, through which nodes can exchange information, and $\mathcal{A}=[a_{ij}]\in \mathbb{R}_+^{M\times M}$ is the nonnegative weighted adjacency matrix, where $a_{ij} > 0$ if $(j,i) \in \mathcal{E}$, and $a_{ij} = 0$ otherwise.
The degree matrix of   $\mathcal{G}$ is  $\mathcal{D}={\rm diag}(d_i)\in \mathbb{R}^{M\times M}$, where $d_i :=\sum_{j=1}^M a_{ij}$, for every $i\in\mathcal{M}$.  The Laplacian matrix associated with the network is  $\mathcal{L} = \mathcal{D} - \mathcal{A}$.   

\begin{assumption} 
	[\emph{Communication network}] 
	The graph $\mathcal{G}$ is undirected and connected.
	\label{Ass:network}
\end{assumption}

\begin{remark}{\bf {\rm \bf(}\emph{Laplacian matrix}{\rm \bf)}} \label{laplacian}
	It follows from Assumption \ref{Ass:network} that the Laplacian $\mathcal L$ associated with the graph is symmetric and irreducible. Therefore, its spectrum is of the form 
	$$
	\sigma(\mathcal L) = \{0,\lambda_2,\dots,\lambda_M\}, \quad {\rm with} \quad 0<\lambda_2\le \dots \le \lambda_M.
	$$
	We also note that $-\mathcal L$ is a compartmental matrix  (see \cite{Haddad2002}), since its off-diagonal entries are nonnegative (and hence it is a Metzler matrix)  and the sum of the entries in each of its columns is nonpositive, i.e., ${\mathbf 1}^\top(-{\mathcal L}) \le {\bf 0}^\top$.  Therefore, $- \mathcal L$ is an irreducible compartmental matrix,  and it satisfies the (stronger) condition ${\mathbf 1}^\top(-\mathcal L) = {\bf 0}^\top$. 
	Given any $i\in {\mathcal M}$,
	if we denote by $\tilde{\mathcal L}$ the matrix obtained from $\mathcal L$ by removing the $i$th row and the $i$th column, namely the entries related to the connections of     node $i$, we have that $\tilde{\mathcal L}  = \tilde{\mathcal L}^\top$, and $-\tilde{\mathcal L}$ is still a compartmental matrix and it is Hurwitz (see Lemma \ref{small_lap} in \ref{sec.App.Laplacian}).
	Consequently, $\tilde{\mathcal L}\succ 0$, and hence $\tilde{\mathcal L} \otimes I\succ 0$. 
\end{remark}

\begin{figure}
		\centering
	\includegraphics[width=0.47\textwidth]{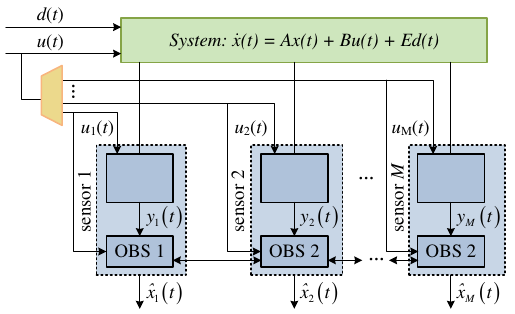}  
	\caption{ Scheme of the proposed distributed sensor network.}
\label{fig.WSN}
	\centering
\end{figure}

We assume that the $i$th sensor node generates the state estimate
at time $t$, $\hat x_i(t)$, through a  DUIO described as follows:
\begin{equation}
\underset{i\in {\mathcal M}}{{\rm DUIO}_i}:
	\left\{\begin{array}{l}
		\begin{aligned}
			\dot z_i(t)
			&=E_i z_i(t)+F_iu_i(t)+L_iy_i(t)\\
			&\quad +K_i \sum\limits_{j=1}^Ma_{ij}[\hat {x}_j(t)-\hat{x}_i(t)] \\
			\hat x_i(t)&=z_i(t)+H_i y_i(t),
		\end{aligned}
	\end{array}\right.
	\label{eq:duio}
\end{equation}
where $z_i(t)\in \mathbb R^{n_x}$ is the  state of the $i$th UIO (${\rm DUIO}_i$), $\hat x_i(t)\in \mathbb R^{n_x}$ is the  estimate of system \eqref{eq:system} state provided by node $i$. The coefficients $a_{ij}$ are the entries of the communication graph adjacency matrix, while $E_i \in \mathbb R^{n_x \times n_x}$, $F_i \in \mathbb R^{n_x \times n_{m_i}}$, $L_i,H_i \in \mathbb R^{n_x \times n_{y_i}}$,  $K_i \in \mathbb{R}^{n_x \times n_x}$ are matrix parameters to be designed. 

In the proposed set-up,  
we provide a first qualitative statement of the estimation problem we address in the paper.

\begin{problem}\label{prob:1}
	Given the   systems ${\mathcal T}_i$, $i \in \mathcal{M}$,  subject to unknown inputs and disturbances, 
	and   communicating through a graph ${\mathcal G}$, satisfying Assumption \ref{Ass:network},
	determine, if possible, the matrices $E_i , F_i , L_i,H_i $, and  $K_i, i
	\in {\mathcal M},$ of 
	the distributed state estimation scheme \eqref{eq:duio}
	in such a way that the state estimates provided by the observers across all nodes
	achieve consensus and the common state estimate converges to the real state value.
\end{problem}

\section{Distributed  Model-based State Estimation}\label{sec.model_based}
If all the system matrices $(A,B,E,C_i)$, $ i\in {\mathcal M},$ are known,  
the problem is analogous to the one investigated  in \cite{yang2022state} and \cite{CaoWang}, where
sufficient conditions for the solvability of the DUIO problem  have been provided by means of  a model-based approach. However, the set-up considered here is a bit more general, since it does not impose any particular structure on the  gain matrices $K_i$ weighting the consensus term (see comments after Lemma \ref{lem.con}).

Upon defining the global estimation error by
concatenation  as $
e_G(t) :=\big[ 
e^\top_1(t) \ \cdots \ e^\top_i(t) \ \cdots \  e^\top_M(t)  \big]^\top$,
where $e_i(t) := x(t) - \hat{x}_i(t)$  is the estimation error of node $i$. Similarly to \cite[Sections 2--3]{valcher1999state} and  \cite[Section 4]{yang2022state}  (see also \cite{CaoWang}), 
taking the time derivative of $e_G(t)$ yields
\begin{align}
	\dot e_G(t) &=  \big[{\rm diag}(E_i)- {\rm diag}(K_i)( \mathcal{L} \otimes I )\big]e_G(t) \nonumber \\
	&+ \big[{\rm diag}(I - H_i 
	C_i){\rm diag}(B_i^m)-  {\rm diag}(F_i)\big]u_G(t) \nonumber \\
	&+    \big[ {\rm diag}(I -  H_iC_i)(I \otimes  A)-{\rm diag}(E_i) \times \nonumber \\
	& \ \quad ~ {\rm diag}(I - H_i 
	C_i)-{\rm diag}(L_i){\rm diag}(C_i)\big]x_G(t) \nonumber \\
	&+ \ {\rm diag}(I -  H_iC_i)    {\rm diag}(B_i^p)w_G(t),   \label{eq:err5}
\end{align}
where $z_G(t)$, $w_G(t)$, and $ u_G(t)$ are  defined analogously to  $e_G(t)$, while $x_G(t):=\big[ 
x^\top(t) \ \cdots \ x^\top(t)  \ \cdots \  x^\top(t)  \big]^\top = {\mathbf 1}\otimes x(t)$.  The derivation of \eqref{eq:err5} is detailed in \ref{sec.App.a}.
Clearly,   the estimation error dynamics is independent of 
the disturbance, the unknown input and the specific input and state trajectories if and only if the following conditions are satisfied
\begin{subequations}\label{eq:equality}
	\begin{align}
		& {\rm diag}(F_i)   = {\rm diag}(I-H_i C_i){\rm diag}(B_i^m), \label{cond.a}\\ 
		&{\rm diag}(I-H_i C_i){\rm diag}(B_i^p)  = {\bf 0}, \label{cond.b}\\
		&{\rm diag}(E_i )   =  {\rm diag}(I- H_i C_i )(I\otimes A)- [{\rm diag}(L_i) \nonumber \\
		&\qquad \qquad \quad -{\rm diag}(E_i){\rm diag}(H_i)]{\rm diag}(C_i).\label{cond.c}
	\end{align}
\end{subequations}
When so, equation \eqref{eq:err5} becomes
\begin{equation}
	\dot e_G(t)= [{\rm diag}(E_i) - {\rm diag}(K_i)( \mathcal{L}  \otimes  I )]e_G(t).   
	\label{eq:errvv}
\end{equation}
To guarantee that the conditions in \eqref{eq:equality} are feasible, we state
the following result, whose proof is  a trivial extension of the single agent case \cite{Darouach,Darouach2} (indeed, all matrices in
\eqref{eq:equality} are block diagonal),
and hence omitted. 
\begin{lemma}[\emph{Solvability of Eqns. \eqref{eq:equality}}]\label{lem.con}
	The following facts are equivalent:
	\begin{itemize}
		\item[i)] Equations \eqref{eq:equality} are simultaneously solvable;
		\item[ii)]  For every $i\in {\mathcal M}$, there exists $H_i\in\mathbb{R}^{n_x\times n_{y_i}}$ such that 
		\begin{equation} \label{eq:hi}
			H_i C_i B_i^p = B_i^p;   
		\end{equation}
		\item[iii)]  For every $i\in {\mathcal M}$, ${\rm rank} (C_i B_i^p)= {\rm rank}(B_i^p) = r_i$.
	\end{itemize}
\end{lemma}

It is worthwhile remarking that the solutions of \eqref{eq:hi} can be parametrized as follows \footnote {We assumed that $B_i^p$ is of full column rank. In case it is not, we can always factorize it as
	$B_i^p= \bar B_i^p D_i$, where $\bar B_i^p$ is of full column rank $\bar r_i := {\rm rank} (B_i^p)$ and $D_i$ is of full row rank $\bar r_i$. 
	When so,  the parametrization can be expressed as follows
	$H_i = \bar B_i^p (C_i \bar B_i^p)^{\dagger} + Y_i [ I- (C_i \bar B_i^p)(C_i \bar B_i^p)^{\dagger}],$ $Y_i \in {\mathbb R}^{n_x \times n_{y_i}}$.}
(see \cite{Darouach}):
\begin{equation}
	H_i =   B_i^p (C_i   B_i^p)^{\dagger} + Y_i [ I- (C_i   B_i^p)(C_i B_i^p)^{\dagger}],
	\label{eq:paramHi}
\end{equation}
where  $Y_i$ is a free parameter. In the sequel, we will refer to the particular solution
$B_i^p (C_i B_i^p)^{\dagger}$ of \eqref{eq:hi} with the symbol $\bar H_i$.

In order to ensure that the estimation error asymptotically converges to zero, we need to also impose that 
${\rm diag}(E_i)- {\rm diag}(K_i)(\mathcal{L}  \otimes  I)$ is  Hurwitz stable.
In  \cite{yang2022state} and \cite{CaoWang}, it has been shown that, under any of the equivalent conditions of Lemma \ref{lem.con}, a sufficient condition for the existence of matrices $K_i, i\in {\mathcal M},$ such that ${\rm diag}(E_i)- {\rm diag}(K_i)(\mathcal{L}  \otimes  I)$ is  Hurwitz stable is that the intersection of the undetectable subspaces of the pairs $((I-\bar H_iC_i)A, C_i), i\in {\mathcal M},$ is the zero subspace. 
When so, a possible choice for the matrices $K_i$ is either $\gamma_i U_i U_i^\top$ (see \cite{CaoWang}), where ${\rm range}(U_i)$ is the undetectable subspace of the pair $((I-\bar H_iC_i)A, C_i)$, or $\gamma_i P_i^{-1}$ (see \cite{yang2022state}), where $P_i$ is a symmetric and positive definite matrix that arises from the solution of a suitable LMI. In both cases, $\gamma_i$ is a positive scalar parameter that drives all the eigenvalues of ${\rm diag}(E_i)- {\rm diag}(K_i)(\mathcal{L}  \otimes  I)$ toward 
the left open half-plane as it grows to $+\infty$.

When addressing the distributed UIO design problem from a data-driven perspective, checking
if the intersection of the undetectable subspaces  is the zero   subspace is not feasible, since the test would be too complicated and not robust to numerical errors.
For this reason we explore a stronger sufficient condition that is easy, as well as robust, to test.

\begin{assumption}[\emph{Detectability of a single node}]
	There exists $i\in {\mathcal M}$ such that the  pair $((I- \bar H_i C_i)A, C_i)$  is detectable. 
	Without loss of generality, we   assume that $i=1$, since we can always relabel the agents  to make this happen. 
	\label{Ass:detect}
\end{assumption}

It is clear that if  ${\rm rank} (C_i B_i^p)= {\rm rank}(B_i^p)$ for every $i\in {\mathcal M}$ and 
Assumption \ref{Ass:detect} holds, 
then the intersection of the undetectable subspaces of the pairs $((I-\bar H_iC_i)A, C_i), i\in {\mathcal M},$ is the zero   subspace, and hence there exist matrices $K_i,i\in {\mathcal M}$, such that ${\rm diag}(E_i)- {\rm diag}(K_i)(\mathcal{L}  \otimes  I)$
is Hurwitz  stable. We  provide in the following an explicit solution to the distributed estimation problem, namely a specific choice of the matrices $E_i , F_i , L_i,H_i $, and  $K_i, i
\in {\mathcal M}$. 

\begin{theorem}{\bf {\rm \bf(}\emph{Construction of a model-based DUIO}{\rm \bf)}} \label{gamma_choice}
	Suppose that Assumptions \ref{Ass:network} and \ref{Ass:detect} (for $i=1$) hold, and   ${\rm rank} (C_i B_i^p)= {\rm rank}(B_i^p)$, for every $i\in {\mathcal M}$. Let $M_1\in {\mathbb R}^{n_x \times n_{y_1}}$ be such that $(I-\bar H_1 C_1) A - M_1 C_1$ is Hurwitz stable.  Set
	\begin{subequations}\label{eq:conditions}
		\begin{align}
			E_1 &=  (I-\bar H_1 C_1) A - M_1 C_1, \label{eq:no1}\\
			E_i &= (I-\bar H_i C_i) A, \qquad i\in \{2,\dots,M\}, \label{eq:no2}\\
			L_1 &= M_1 + E_1 \bar H_1, \label{eq:no3}\\
			L_i &= E_i \bar H_i, \qquad \qquad \quad i\in \{2,\dots,M\}, \label{eq:no4}\\
			F_i &= (I - \bar H_iC_i) B_i^m, \quad  i\in {\mathcal M}, \label{eq:no5}\\
			K_1 &= 0, \label{eq:no6}\\
			K_i &= \gamma I, \qquad \qquad  \qquad i\in \{2,\dots,M\}, \label{eq:no7}
		\end{align}
	\end{subequations}
	with
	\begin{equation}
		\gamma > \frac{\Vert \tilde E + {\tilde E}^\top\Vert}{2\lambda_{min}(\tilde{\mathcal L} \otimes I)},
		\label{eq:gammamin}
	\end{equation}
	where $\lambda_{min}(\tilde{\mathcal L} \otimes I)>0$ is the smallest eigenvalue of $\tilde{\mathcal L} \otimes I \succ 0$ (see Remark \ref{laplacian}), and 
	$\tilde E := {\rm diag}(E_i)_{i=2}^M.$
	Then, for this choice of the matrices, the model-based DUIO in \eqref{eq:duio} can reconstruct the system state asymptotically. 
\end{theorem}

\begin{remark}
	Note that the procedure proposed here leads to a rather simplified solution since it suffices to stabilize the leader agent and then choose a suitable $\gamma$ to stabilize also the other agents. Therefore, the matrices $E_i$ for $i \in\{2,\dots,M\}$ play no role, except in setting the bound on $\gamma$, which is the reason why we imposed conditions \eqref{eq:no2}. 
\end{remark}
						
\section{Distributed Data-driven  State Estimation}\label{sec.DUIO}
	Throughout this section, we make the following  additional assumption on the systems ${\mathcal T}_i, i\in {\mathcal M}$.
	\begin{assumption}{\bf{\rm\bf(}\emph{Unknown system model}{\rm \bf)}} 
		For each $i\in \mathcal M$, the matrices $(A,B,E,C_i)$ of the system ${\mathcal T}_i$ are unknown.
		\label{Ass:system}
	\end{assumption}
	
	Under the previous Assumption, the model-based  method for designing   DUIOs   described in the previous section does not apply anymore. Hence, to solve Problem \ref{prob:1}, 
	we explore the possibility of designing a  DUIO based on data. 
	
	In industrial processes, it  is not always  feasible or safe to measure real-time states and transmit them to remote sensors.  However,  offline experiments can be conducted to gather state data, which are further sent to remote sensors
	for the design of state observers that can operate online.
	In line with recent studies \cite{berberich2020data,berberich2020robust,turan2021data,liu2022data1}, we assume that input/output/state data can be collected and make the
	following assumption.

	\begin{assumption}{\bf {\rm \bf(}\emph{Offline and online data acquisition}{\rm \bf)}}\label{Ass:offline} 
		During the offline phase, input/output/state sampled data $\bar{u}_i:=\{\bar{u}_i({ t_k})\}_{k=0}^{N-1}$, $\bar{y}_i:=\{\bar{y}_i({ t_k})\}_{k=0}^{N-1}$, $\dot {\bar{y}}_i:=\{\dot{\bar{y}}_i({ t_k})\}_{k=0}^{N-1}$, $\bar{x}_i:=\{\bar{x}_i({ t_k})\}_{k=0}^{N-1}$ and  $\dot{\bar{x}}_i:=\{\dot {\bar{x}}_i({ t_k})\}_{k=0}^{N-1}$ are collected locally by each system ${\mathcal T}_i$  described as in \eqref{eq:system}--\eqref{eq:output}, 
		possibly corresponding to different initial states and inputs\footnote{ No special requirement is imposed on the  sampling times $\{t_k\}_{k=0}^{N-1}$. Indeed, each agent $i$ could have 
			different sampling times, say $\{t_k^i\}_{k=0}^{N-1}$.}.
		During  online operation, only the inputs $u_i(t)$, { outputs $y_i(t),$ and the output derivatives} $\dot y_i(t), i\in {\mathcal M}$,  { $t\in {\mathbb R}_+$}, are available. 
	\end{assumption}
	
	\begin{remark} 
		Assumption \ref{Ass:offline}  indicates that although the matrices $(A,B,E,C_i)$ are unknown, the measurable input and 
		output data are available both offline and online. On the other hand, states can only be obtained through offline experiments, but are not available during online operations. This setting can be fulfilled in a remote control scenario, and it has been widely considered in the data-driven state-estimation literature, see e.g., \cite{turan2021data,TAC-UIO, liu2023learning,MHE}. 
		
		When the state and the output derivatives are not physical quantities, their computation is likely to be error-prone. However, after having recorded the state and output trajectories with a high sampling rate, we can obtain a good approximation of their values in a post-processing step, since the data are collected in an offline phase. We can explicitly account for errors in the computation of the derivatives by modeling these errors as a measurement noise (see \cite{BERBERICH2021}). Alternatively, when the derivatives are difficult to compute, we can resort to an integral version of the relation in \eqref{eq:systemi} and \eqref{eq:output}, which leads to an equivalent characterization, as it is shown in \cite{eventTrig}. For the sake of simplicity, we carry on the analysis using the derivatives. However, all the results can be adapted with no further effort to use the integral representation of the data.  
	\end{remark}
	
	Under Assumption \ref{Ass:offline}, we define for every $i\in {\mathcal M}$ the following matrices:   
	\begin{align}\label{eq:pf}
		{U}_i \! &:= \!  \begin{bmatrix}
			\bar u_i^\top(t_0) \ \cdots  \ \bar u_i^\top(t_{N-1}) \\
		\end{bmatrix}^\top \! \in {\mathbb R}^{{n_{m_i}} \! \times \!  N} \nonumber \\
		{Y}_i\! &:= \! \begin{bmatrix}
			\bar y_i^\top(t_0) \ \cdots \ \bar y_i^\top(t_{N-1})  \\
		\end{bmatrix}^\top \! \in {\mathbb R}^{{n_{y_i}} \! \times \!  N} \nonumber \\
		\dot{Y}_i \! &:= \! \begin{bmatrix}
			\dot {\bar y}_i^\top(t_0) \ \cdots \  \dot {\bar y}_i^\top(t_{N-1})  \\
		\end{bmatrix}^\top \! \in {\mathbb R}^{{n_{y_i}} \! \times \!  N} \nonumber \\	
		{X}_i \!&:=  \! \begin{bmatrix}
			\bar x_i^\top(t_0) \ \cdots \  \bar x_i^\top(t_{N-1})  \\
		\end{bmatrix}^\top \! \in {\mathbb R}^{{n_x} \! \times \!  N} \nonumber \\
		\dot{X}_i \! &:= \! \begin{bmatrix}
			\dot {\bar x}_i^\top(t_0) \ \cdots \ \dot {\bar x}_i^\top(t_{N-1})  \\
		\end{bmatrix}^\top \! \in {\mathbb R}^{{n_x} \! \times \!  N}.
	\end{align}
In addition,  even if we cannot measure the unknown input $w_i$, it is convenient to introduce the sequence $\bar{w}_i:=\{\bar{w}_i(t_k)\}_{k=0}^{N-1}$   and the corresponding matrix 
$$W_i:= \begin{bmatrix}
	\bar w_i^\top(t_0) \ \cdots  \ \bar w_i^\top(t_{N-1})  \\
\end{bmatrix}^\top \in {\mathbb R}^{{r_i} \times  N}.$$ 
We make the following assumption on the pre-collected data. 

\begin{assumption} {\bf {\rm \bf(}\emph{Rank of pre-collected data}{\rm \bf)}} For each $i \in \mathcal{M}$, it holds that 
	$${\rm rank}\left(\begin{bmatrix}
		U_i^\top  \ W_i^\top  \ X_i^\top   \end{bmatrix}^\top \right) =n_{m_i}+ r_i +n_x.$$
	\label{Ass:hisiuxy}
\end{assumption}

\begin{remark}{\bf {\rm \bf(}\emph{Conservativeness of Assumption \ref{Ass:hisiuxy}}{\rm \bf)}}
	\label{rem.resrict}
	As it will be proved in Theorem \ref{The.1}, Assumption \ref{Ass:hisiuxy} ensures that any input/output/state trajectory of system ${\mathcal T}_i$ can be represented as a linear combination of 
	the columns of
	$[U_i^\top  \ X_i^\top \ Y_i^\top ]^\top$.
	In the case of  discrete-time systems,  according to \emph{Willems et al.}'s fundamental lemma \cite{willems2005note}, Assumption \ref{Ass:hisiuxy} is fulfilled when the pair $(A,[B\ E])$ is controllable and the input and disturbance signal  
	$\{[u_i^\top(t_k) \ w_i^\top(t_k)]^\top \}_{k=0}^{N-1}$ is persistently exciting of order $n_x+2$ (see \cite{turan2021data,TAC-UIO}).
	The relationship between 
	Assumption \ref{Ass:hisiuxy} and persistence of excitation is more involved for  continuous-time systems, and we refer the interested reader to \cite{contWillems}. 
	However, since no constraint is imposed on the sampling times $\{t_k\}_{k=0}^{N-1}$, it is easy to conceive experiments in such a way that the collected data satisfy   Assumption \ref{Ass:hisiuxy}.
	Finally,  it is worth emphasizing that   Assumption \ref{Ass:hisiuxy} does not allow system identification from the collected data. Specifically,  the presence of unmeasured disturbances $w_i(t)$ in  ${\mathcal T}_i$ 
	does not guarantee  the possibility of identifying $A, B_i^m$ and  $B_i^p$  from data.
\end{remark}

In the rest of the paper we will focus on this revised version of Problem \ref{prob:1}.

\begin{problem}\label{prob:2}
	Given the unknown systems ${\mathcal T}_i$, $i \in \mathcal{M}$, subject to unknown inputs and disturbances, and satisfying Assumptions 
	\ref{Ass:network}--\ref{Ass:hisiuxy}, design, if possible, a distributed state estimation scheme described as in \eqref{eq:duio},
	whose matrices are derived from the offline data,
	such that the state estimates provided by the observers across all nodes
	achieve consensus and the common state estimate converges to  the real state value.
\end{problem}

To address Problem \ref{prob:2}, we build upon the data-driven UIO for a single agent  
proposed in \cite[Section \uppercase\expandafter{\romannumeral2}]{turan2021data}  in a discrete-time setting and develop its distributed version for  continuous-time   systems.  
The main ideas behind the data-driven UIO proposed in \cite{turan2021data} are substantially two.
First of all, a  UIO for a single sensor ${\mathcal T}_i$,  described as in \eqref{eq:duio},
includes among its input/output trajectories all the 
(control) input/output/state trajectories of system ${\mathcal T}_i$ (see \cite[Remark 3]{turan2021data}).
Secondly, if the  historical data are sufficiently rich to capture the dynamics of 
the (control) input/output/state trajectories of system ${\mathcal T}_i$, then they can be used to design the matrices of the $i$th UIO (${\rm DUIO}_i$).
We will follow a similar path and first derive (see Theorem \ref{The.1}) conditions that ensure that   historical data allow to identify the online trajectories of ${\mathcal T}_i$. 
Subsequently, in Theorem \ref{DDexistenceDDUIO}, we   propose a sufficient condition for  the existence of a data-driven DUIO, namely a DUIO described as in \eqref{eq:duio}, whose matrices are obtained from the historical data. The explicit expression of these matrices as well as a possible parametrization of them  is given in Theorem \ref{The.2}.

\subsection{Consistency of Offline and Online Trajectories}\label{sec.3.1}
Inspired by \cite[Lemma 1]{turan2021data} (see also Lemma 7 in \cite{TAC-UIO}), in Theorem  \ref{The.1} we establish the consistency of offline and online trajectories affected by unknown inputs and noise. We first recall the concept of data compatibility \cite[Definition 5]{turan2021data}, which determines whether the  historical data $(\bar u_i,\bar y_i, \dot {\bar y}_i,\bar x_i, \dot {\bar x}_i)$ are sufficiently representative of system ${\mathcal T}_i$ trajectories.

\begin{definition}{\bf {\rm \bf(}\emph{Data compatibility}{\rm \bf)}}\label{def.compa}
	An input/output/ \\
	state trajectory $(\{ u_i(t) \}_{{ t\in {\mathbb R}_+}},\{ y_i(t)\}_{{ t\in {\mathbb R}_+}}, \{\dot y_i(t) \}_{{ t\in {\mathbb R}_+}},$ 
	$\{x(t) \}_{{ t\in {\mathbb R}_+}}, \{\dot x(t) \}_{{ t\in {\mathbb R}_+}})$ is compatible with the  historical data $(\bar u_i,\bar y_i,\dot {\bar y}_i, \bar x_i, \dot {\bar x}_i)$  if the following condition holds
	\begin{equation}
		\left[ \begin{matrix}
			u_i(t)  \\
			y_i(t)  \\
			\dot {y}_i(t)  \\
			x(t)  \\
			\dot x(t)  \\
		\end{matrix} \right]\!\in\! {\rm range}\! \left( \left[ \begin{matrix}
			U_i  \\
			Y_i  \\
			{ \dot{Y}_i}  \\
			X_i  \\
			{ \dot{X}_i}  \\
		\end{matrix} \right] \right), ~~ \forall { t\in {\mathbb R}_+},
		\label{eq:range}\end{equation}
	where $U_i$, $Y_i$,  
	${\dot{Y}_i}$, $X_i$, and ${\dot{X}_i}$ are defined in \eqref{eq:pf}.   The set of trajectories compatible with the  historical data $(\bar{u}_i, \bar{y}_i,\dot {\bar y}_i, \bar{x}_i,\dot {\bar x}_i)$ is defined as
	\begin{align}
		&T(\bar{u}_i,\bar{y}_i, \dot{\bar{y}}_i,\bar{x}_i,\dot {\bar x}_i) 
		:=  \big\{\big(\{u_i(t)\}_{{ t\in {\mathbb R}_+}},\{y_i(t)\}_{{ t\in {\mathbb R}_+}}, \nonumber\\
		&\quad \quad \{\dot {y}_i(t)\}_{{ t\in {\mathbb R}_+}}, \{ x(t) \}_{{ t\in {\mathbb R}_+}},\{ \dot x(t) \}_{{ t\in {\mathbb R}_+}}\big)\ \big|\ \eqref{eq:range} \ \text{holds} \big\}.	\label{eq:hisandtra}
	\end{align}
\end{definition}
The set of input/output/state trajectories $(\{u_i(t)\}_{{ t\in {\mathbb R}_+}},$ $\{ y_i(t)\}_{{ t\in {\mathbb R}_+}}, \{ \dot {y}_i(t)\}_{{ t\in {\mathbb R}_+}},\{x(t)\}_{{ t\in {\mathbb R}_+}}, \{\dot x(t)\}_{{t\in {\mathbb R}_+}})$ compatible with the equations of system ${\mathcal T}_i$ is defined as follows
\begin{align}
	T_{{\mathcal T}_i}& \! := \! \big\{ 
	\big(\{u_i(t)\}_{{ t\in {\mathbb R}_+}},\{ y_i(t)\}_{{ t\in {\mathbb R}_+}},\{\dot {y}_i(t)\}_{{ t\in {\mathbb R}_+}},\{x(t)\}_{{ t\in {\mathbb R}_+}},  \nonumber \\
	&\quad \quad ~  \{\dot x(t)\}_{{ t\in {\mathbb R}_+}}\big)\big| \exists\{ w_i(t) \}_{{ t\in {\mathbb R}_+}}  \ \text{such that} \ \big(\{u_i(t)\}_{{ t\in {\mathbb R}_+}}, \nonumber\\
	&\quad \quad ~  \{ y_i(t)\}_{{ t\in {\mathbb R}_+}},  \{\dot {y}_i(t)\}_{{ t\in {\mathbb R}_+}}, \{x(t)\}_{{ t\in {\mathbb R}_+}},  \{\dot x(t)\}_{{ t\in {\mathbb R}_+}} \nonumber\\
	&\quad \quad ~
	\{w_i(t)\}_{{ t\in {\mathbb R}_+}} \big)   \ \text{satisfies}\ \eqref{eq:systemi} \text{-} \eqref{eq:output} \big\} .\label{eq:trand1} 
\end{align}

\begin{theorem}{\bf {\rm \bf(}\emph{Consistency of offline and online trajectories}{\rm \bf)}}\label{The.1}
	Under Assumptions 
	\ref{Ass:system}--\ref{Ass:hisiuxy}, 
	for every $i\in \mathcal{M}$,   the set of trajectories compatible with the  offline data $(\bar{u}_i,\bar{y}_i, \dot {\bar{y}}_i,\bar{x}_i, \dot {\bar{x}}_i)$  coincides with the set of trajectories of the  system ${\mathcal T}_i$, i.e.,  $T(\bar{u}_i,\bar{y}_i, \dot {\bar{y}}_i, \bar{x}_i,\dot {\bar{x}}_i)=T_{{\mathcal T}_i}$.
\end{theorem}
	
\subsection{Existence and construction of a D-DUIO}\label{sec.3.2}
To extend the analysis carried on in Section \ref{sec.model_based}, under Assumption \ref{Ass:detect}, we first need to understand how one can deduce from data the existence of an agent for which Assumption \ref{Ass:detect} holds and the fact that for every agent one of the equivalent conditions of Lemma \ref{lem.con} holds.
We introduce the following technical result.

\begin{lemma}{\bf {\rm \bf(}\emph{Solvability conditions in terms of collected data}{\rm \bf)}} \label{Lemma.DD}
Suppose that   Assumptions 
\ref{Ass:system}--\ref{Ass:hisiuxy} hold. 
Then  $\forall\ i\in {\mathcal M}$
\begin{itemize}
\item[i)] Condition 
${\rm rank} (C_i B_i^p)= {\rm rank}(B_i^p)$  (i.e., condition {\em iii)} of Lemma \ref{lem.con}) holds if and only if
\begin{equation}
	{\rm rank} \left(\begin{bmatrix} U_i\cr
		\dot Y_i\cr
		X_i\end{bmatrix} \right) = {\rm rank} \left(\begin{bmatrix} U_i\cr
		X_i\cr
		\dot{X}_i\end{bmatrix} \right).
	\label{eq:2ranks}
\end{equation}
\item[ii)] Assumption \ref{Ass:detect} holds if and only if  there exists $i\in {\mathcal M}$ such that
\begin{equation}
	\label{eq:DDdetecti}
	{\rm rank}\begin{bmatrix}
		sX_i- \dot{X}_i \\
		U_i \\
		Y_i
	\end{bmatrix} = n_x+n_{m_i}+ r_i,  \forall s\in \mathbb C, {\rm Re}(s) \ge 0.
\end{equation}
\end{itemize}
\end{lemma}

As an immediate consequence of Lemma \ref{Lemma.DD} and of Theorem \ref{gamma_choice}, we can claim what follows.

\begin{theorem}{\bf {\rm \bf(}\emph{Existence of D-DUIO}{\rm \bf)}}\label{DDexistenceDDUIO}
If   the communication graph  ${\mathcal G}$ is undirected and connected (namely, Assumption \ref{Ass:network} holds),  condition 
\eqref{eq:DDdetecti} holds for one   index $i\in {\mathcal M}$ (namely, Assumption \ref{Ass:detect} holds), say $i=1$,
and    condition 
\eqref{eq:2ranks} holds for every $i\in {\mathcal M}$, then there exists a  distributed UIO in \eqref{eq:duio} that asymptotically estimates the  state of the original system. 
\end{theorem}

The previous result provides    a way to check on data
the  existence of an asymptotic D-DUIO. We want now to enable the explicit construction of the matrices of such an asymptotic D-DUIO using only the collected data. \\
To this end, we preliminarily notice that
each $X_i$ is of full row rank, as a result of Assumption \ref{Ass:hisiuxy}, and hence from
$Y_i = C_i X_i$, we can immediately deduce $C_i$ for every $i\in {\mathcal M}$ as
$C_i = Y_i X_i^\dagger.$

We are now ready to prove our main result. 

\begin{theorem}{\bf {\rm \bf(}\emph{Construction of D-DUIO}{\rm \bf)}}\label{The.2}
Suppose that  the communication graph  ${\mathcal G}$ is undirected and connected (namely, Assumption \ref{Ass:network} holds), that condition 
\eqref{eq:DDdetecti} holds for a single index $i\in {\mathcal M}$ (namely, Assumption \ref{Ass:detect} holds), say $i=1$,
and that  condition 
\eqref{eq:2ranks} holds for every $i\in {\mathcal M}$. Then:
\begin{itemize}
\item[i)] For every $i\in {\mathcal M}$ there exist matrices $T_u^i, T_y^i,$ and $T_x^i$, of suitable dimensions, with 
${\rm rank} (T_y^i)= r_i$, such that
\begin{equation}
\dot{X} _i = \begin{bmatrix}
T_u^i & T_y^i &T_x^i \end{bmatrix} \begin{bmatrix} U_i\cr \dot{Y}_i \cr X_i\end{bmatrix}. 
\label{eq:mainrel}
\end{equation}
Moreover, for $i=1$ the pair $(T_x^i,C_i)$ is detectable.
\item[ii)] For every solution $T_u^1, T_y^1,$ and $T_x^1$ of \eqref{eq:mainrel} for which 
${\rm rank} (T_y^1)= r_1$, the pair  $(T_x^1,C_1)$ is detectable.
\item[iii)]  Let $M_1$ be a matrix such that $T_x^1 - M_1 C_1$ is Hurwitz stable.
If we assume
\begin{subequations}\label{eq:conditionsDD}
\begin{align}
E_1 &=  T_x^1  - M_1 C_1, \label{eq:DD1}\\
E_i &= T_x^i, \qquad i\in \{2,\dots,M\}, \label{eq:DD2}\\
L_1  &  = M_1 + E_1 T_y^1 , \label{eq:DD3}\\
L_i &  = E_i  T_y^i, \quad i\in \{2,\dots,M\}, \label{eq:DD4}\\
F_i &= T_u^i, \qquad i\in {\mathcal M}, \label{eq:DD5}\\
H_i &= T_y^i, \qquad i\in {\mathcal M}, \label{eq:DD6}\\
K_1 &= 0, \label{eq:DD7}\\
K_i &= \gamma I, \qquad i\in \{2,\dots,M\}, \label{eq:DD8}
\end{align}
\end{subequations}
and $\gamma$ is chosen such that 
\eqref{eq:gammamin} holds,
where
$\tilde E := {\rm diag}(E_i)_{i=2}^M$,
then the  distributed UIO in \eqref{eq:duio}, for this choice of the matrices, can reconstruct the system state asymptotically. 
\end{itemize}
\end{theorem}

\section{Simulation Results}\label{sec.sim}
This section 
presents a  numerical example   to illustrate the performance of the proposed D-DUIO.
The performance of  this  method is compared to those of a system identification-based (ID) DUIO, as well as of a DUIO based on the exact system model.
The comparison 
demonstrates the effectiveness of the D-DUIO.

\subsection{Performance of D-DUIO}\label{sec.sim1}
\begin{figure}[t]
\centering
\includegraphics[width=0.46\textwidth]{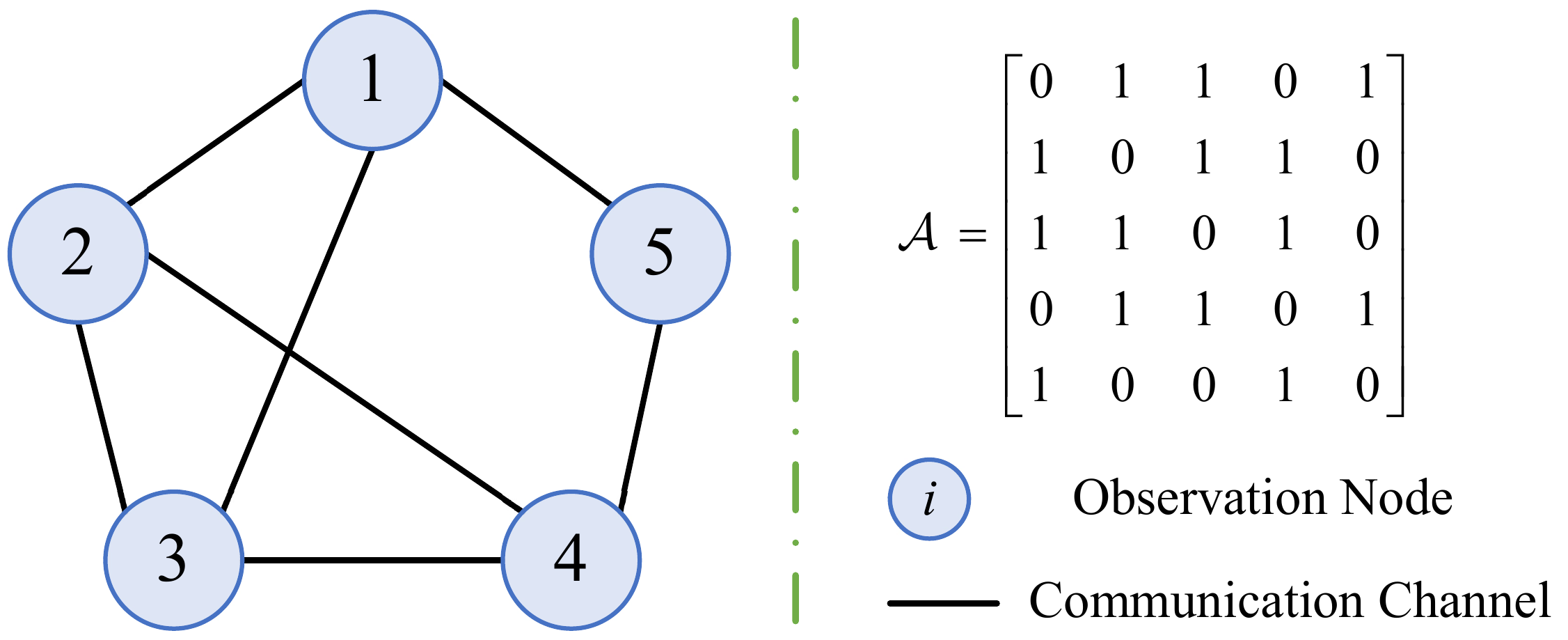}  
\caption{The sensor network topology.}
\label{fig.topo}
\end{figure}

Consider a two-mass-spring system with external disturbances \cite{scis2023liyifei} represented by \eqref{eq:system}--\eqref{eq:output} and a wireless sensor network consisting of $M=5$  nodes.  The undirected and connected communication graph $\mathcal{G}$ is shown in Fig. \ref{fig.topo}.  The system matrices are given by
\begin{align}\label{eq:syssim}
&A=\left[ \begin{matrix}
0 & 1 & 0  & 0 \\
-5.3333 & 0 & 2.6667  & 0 \\
0 & 0 & 0  & 1 \\
2.6667 & 0 & -2.6667 &  0 \\
\end{matrix} \right], \quad 
E=\left[ \begin{matrix}
	0.1  \\  0   \\ 0.1  \\  0
\end{matrix} \right],   \\
&B^m_i=\left[ \begin{matrix}
0  \  1.3333   \ 0  \  0
\end{matrix} \right]^\top ~  i\in  {\mathcal M},\ \quad 
B^u_1=\left[ \begin{matrix}
1 \ \ 1 \  \ 1 \ \ 1  
\end{matrix} \right]^\top\!\!, \nonumber \\
&   B^u_2=\left[ \begin{matrix}
0.5  \ 0.5    \ 0.5  \ 0.5  
\end{matrix} \right]^\top\!\!,  
B^u_3=\left[ \begin{matrix}
0.33 \  0.33   \ 0.33  \ 0.33  
\end{matrix} \right]^\top\!\!, \nonumber \\
& B^u_4=\left[ \begin{matrix}
0.25 \  0.25    \ 0.25  \ 0.25 
\end{matrix} \right]^\top\!\!, B^u_5=\left[ \begin{matrix}
0.2  \ 0.2    \ 0.2  \ 0.2  
\end{matrix} \right]^\top\!\!, \nonumber  
\end{align}
with unknown inputs $u_i^u(t)=(0.2 \times i)\cos(0.2 \times t+2)$, $i\in {\mathcal M}$, and process noise $d(t)$ randomly generated from $[-0.1,0.1]$. 
For every $i\in {\mathcal M}$, we assume that the known inputs are generated by the autonomous system $u_i(t+1)=A_uu_i(t)$, with $A_u=0.5$ and initial condition $u_i(0)$ whose entries are randomly generated  in $[0,1]$.
The outputs of the target system are observed using five nodes whose matrices are  given respectively by
\begin{align}\label{eq:measim}
&C_1=\left[ \begin{matrix}
1 & 0 & 1 & 0  \\
0 & 1 & 0 & 0  \\
0 & 0 & 1 & 1  \\
0 & 0 & 0 & 1  \\
\end{matrix} \right],
C_2=\left[ \begin{matrix}
0 & 1 & 0 & 0  \\
1 & 0 & 1 & 0  \\
0 & 0 & 0 & 1  \\
0 & 0 & 1 & 1  \\
\end{matrix} \right], 
C_3=\left[ \begin{matrix}
	0 & 0 & 1 & 1  \\
	0 & 1 & 0 & 0  \\
	1 & 0 & 1 & 0  \\
	0 & 1 & 1 & 0  \\
\end{matrix} \right], \nonumber \\
&C_4=\left[ \begin{matrix}
1 & 0 & 1 & 1  \\
0 & 1 & 0 & 0  \\
0 & 0 & 1 & 1  \\
1 & 0 & 1 & 0  \\
\end{matrix} \right],  
C_5=\left[ \begin{matrix}
1 & 0 & 1 & 0  \\
0 & 0 & 1 & 1  \\
0 & 0 & 1 & 1  \\
0 & 0 & 0 & 1  \\
\end{matrix} \right]. 
\end{align}

The noisy historical input/output/state trajectories at each node are collected from the linear system \eqref{eq:syssim}--\eqref{eq:measim} with a random initial state. We assume to collect  $N=50$ samples. Moreover, following Theorem \ref{gamma_choice}, the parameter $\gamma$ of the D-DUIO is set to $5$.

\begin{figure}[h]
\includegraphics[width=0.47\textwidth]{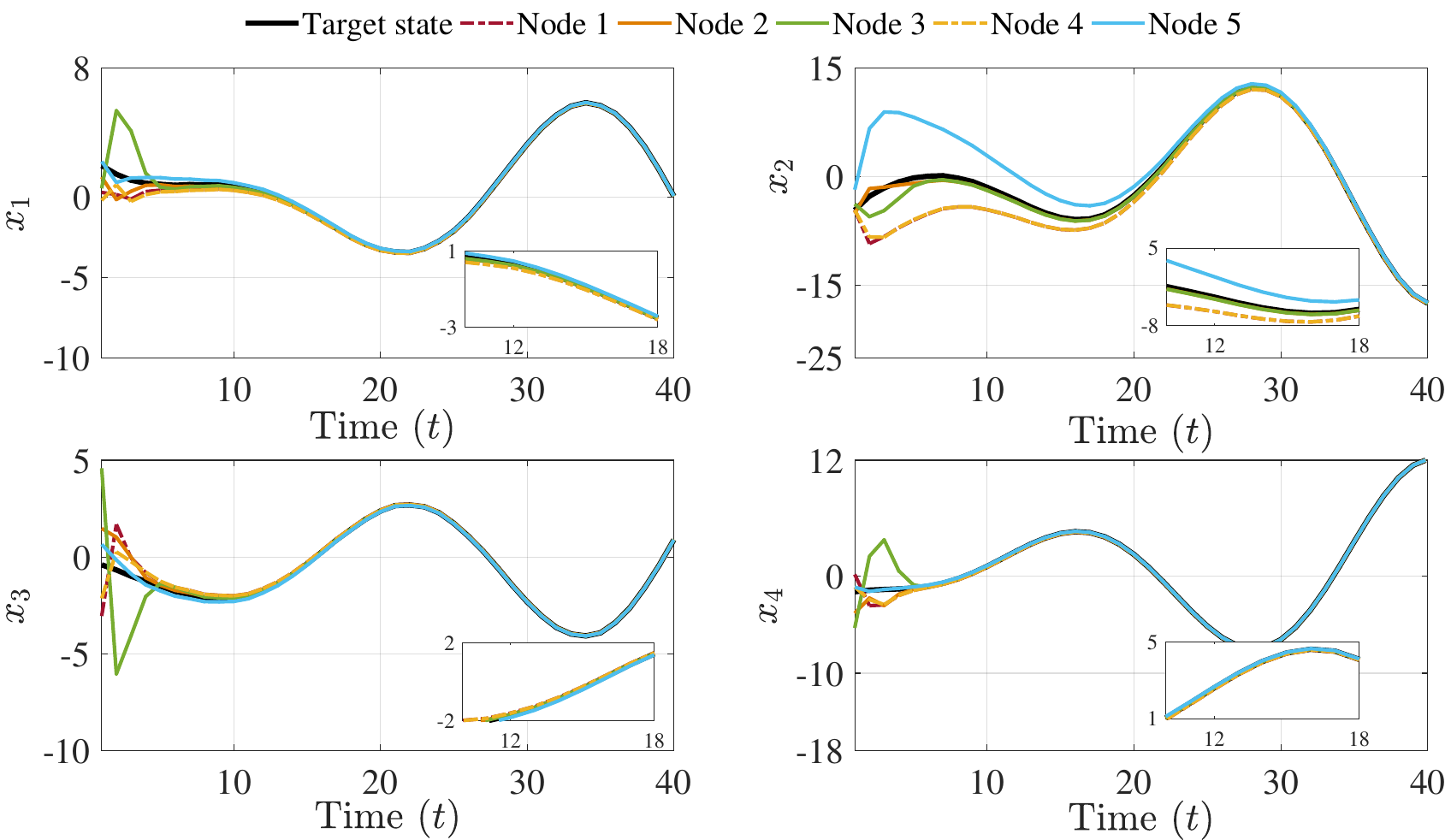}  
\caption{The estimation performance of D-DUIO.}
\label{fig.D-DUIO}
\end{figure}

\begin{figure}[h]	
\includegraphics[width=0.47\textwidth]{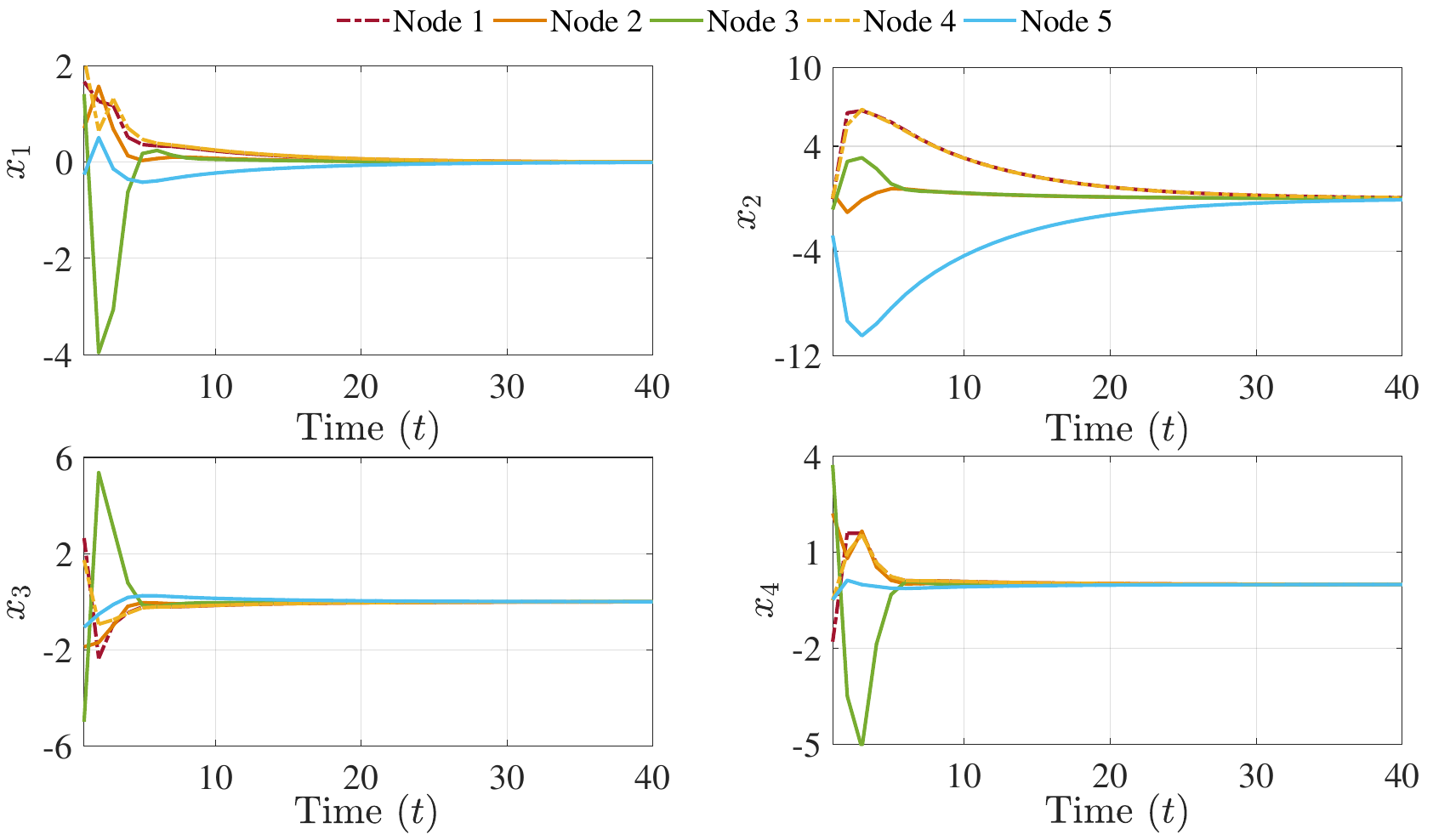}  
\caption{The estimation error of D-DUIO.}
\label{fig.error}
\end{figure}

We present the estimation performance of the D-DUIO in Fig. \ref{fig.D-DUIO}, which shows the state estimates obtained over the simulation  window $[0,40]$. The plots indicate that all nodes achieve consensus on state estimates.  Moreover, the state estimation errors shown in Fig. \ref{fig.error} converge to $0$ asymptotically. This demonstrates that inaccurate estimation arising from unknown inputs and disturbances can be  overcome by the proposed D-DUIO.

\subsection{Comparison with other DSE Methods}\label{sec.sim2}
The proposed  D-DUIO as well as two other DSE methods, namely model-based DUIO and ID-DUIO approaches, are numerically compared in this section.

For ID-DUIO, the coupling matrices $(B^u_i,E)$ of unknown inputs and noises are assumed  known, and the system matrices $(A,B^m_i,C_i)$ are identified by the least-squares method using the same set of offline data. The matrices $(E_i,F_i,L_i,H_i)$ of DUIO and ID-DUIO are obtained by solving equations \eqref{eq:equality}, computed using the CVX toolbox \cite{cvx}.  Parameter $\gamma$ of the two methods is set to $5$.

\begin{figure}[h]
\includegraphics[width=0.47\textwidth]{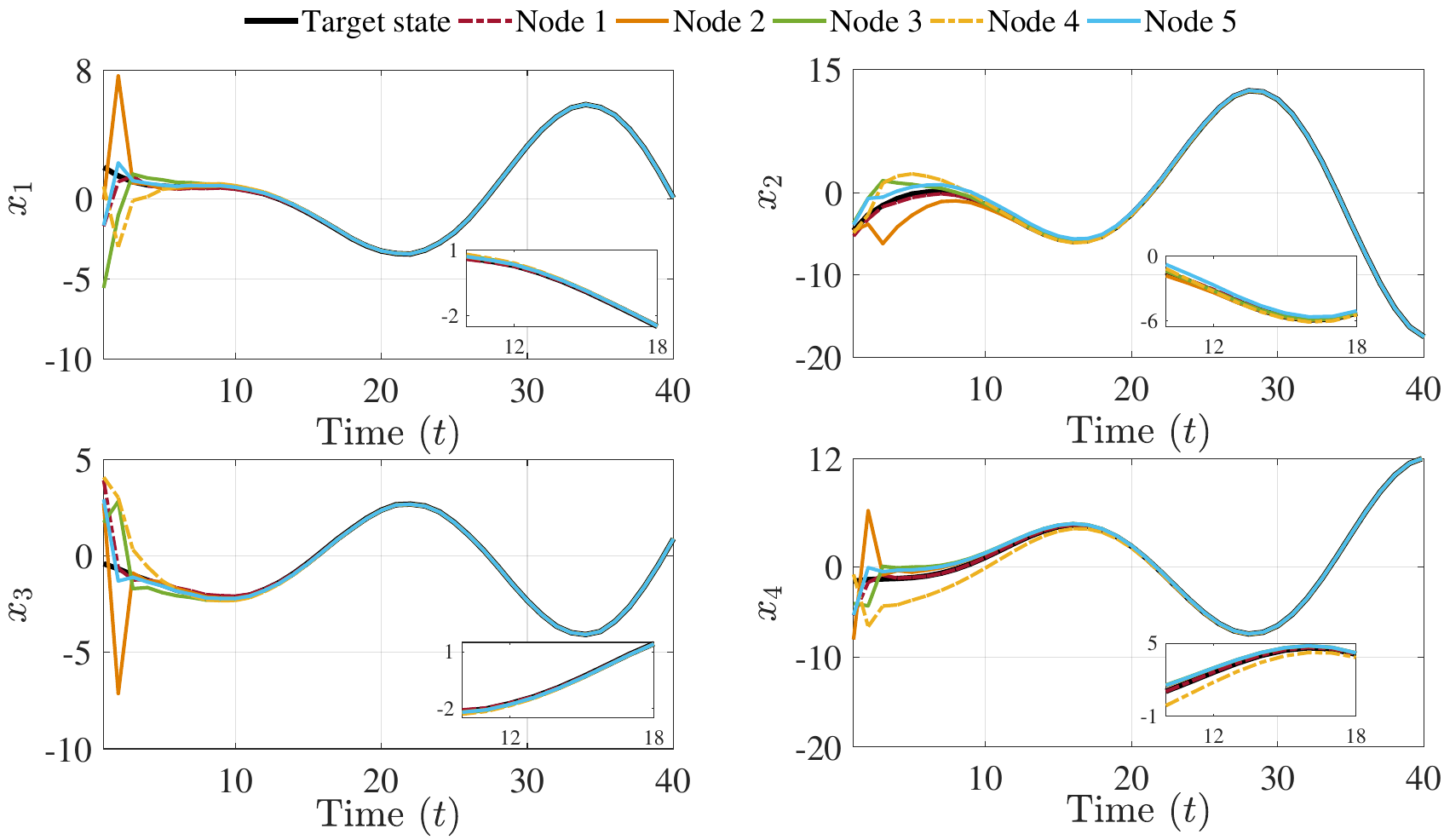}  
\caption{The estimation performance of DUIO.}
\label{fig.DUIO}
\end{figure}

\begin{figure}[h]
\includegraphics[width=0.47\textwidth]{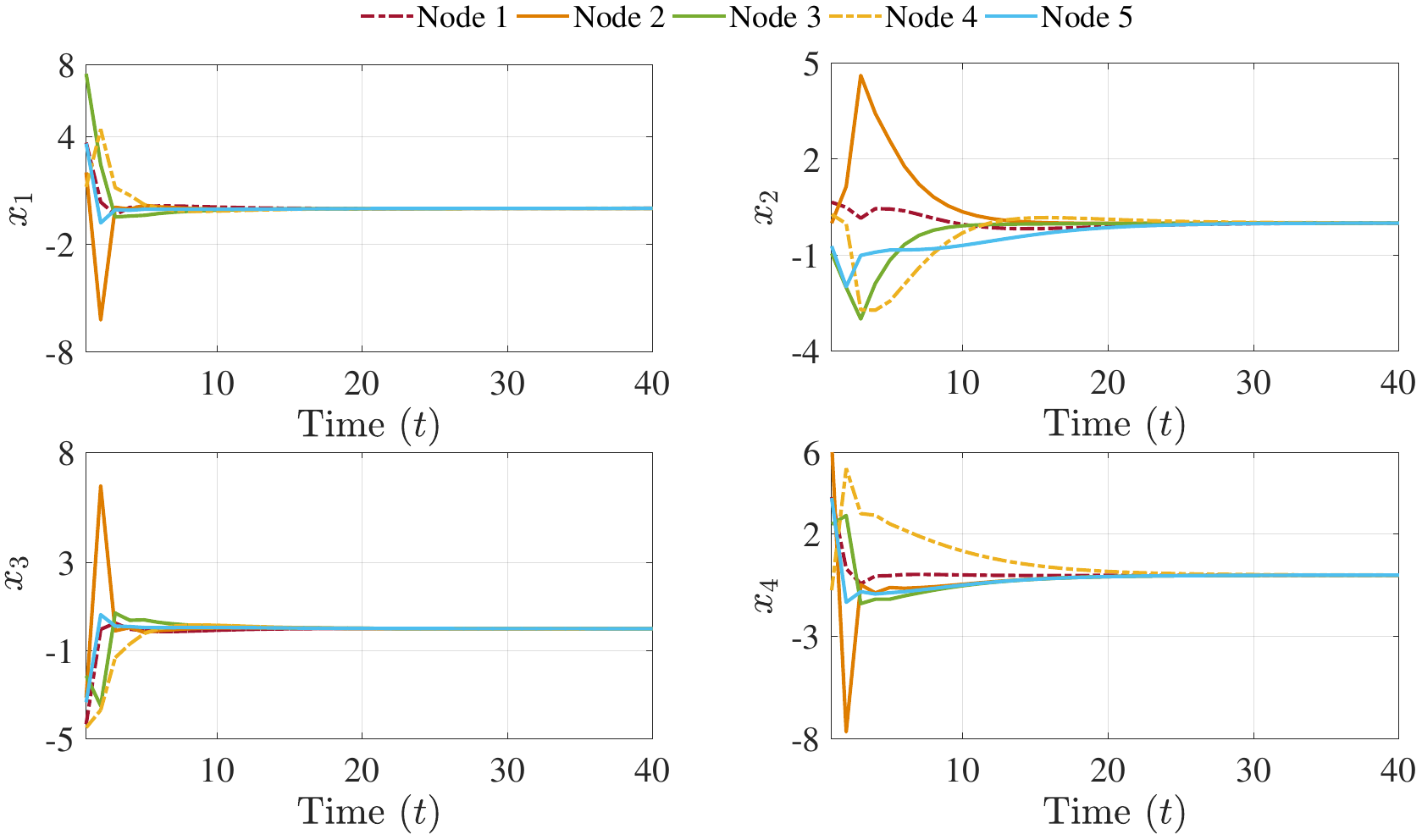}  
\caption{The estimation error of DUIO.}
\label{fig.errDUIO}
\end{figure}

\begin{figure}[h]
\includegraphics[width=0.47\textwidth]{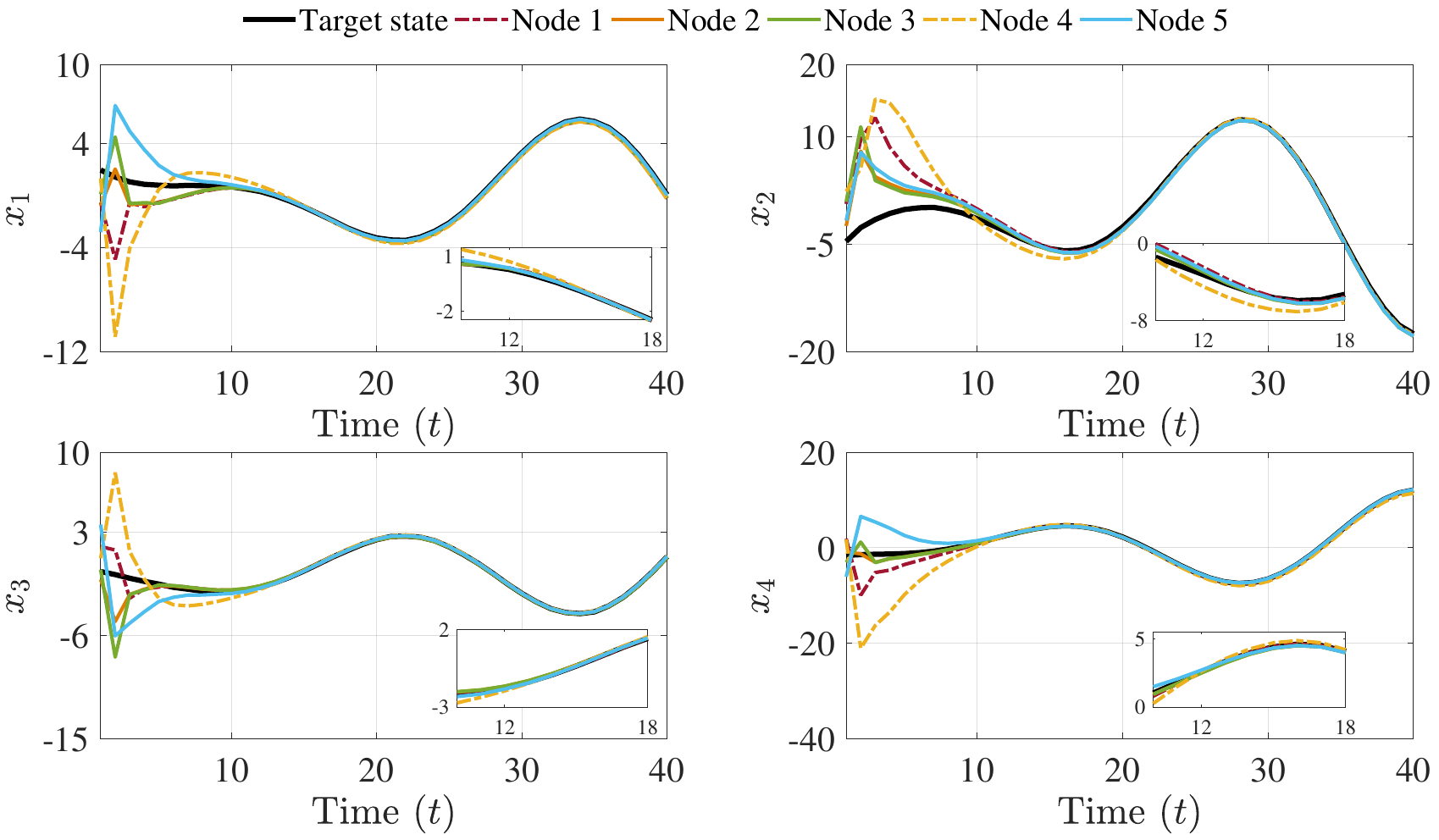}  
\caption{The estimation performance of ID--DUIO.}
\label{fig.ID-DUIO}
\end{figure}

\begin{figure}[h]
\includegraphics[width=0.47\textwidth]{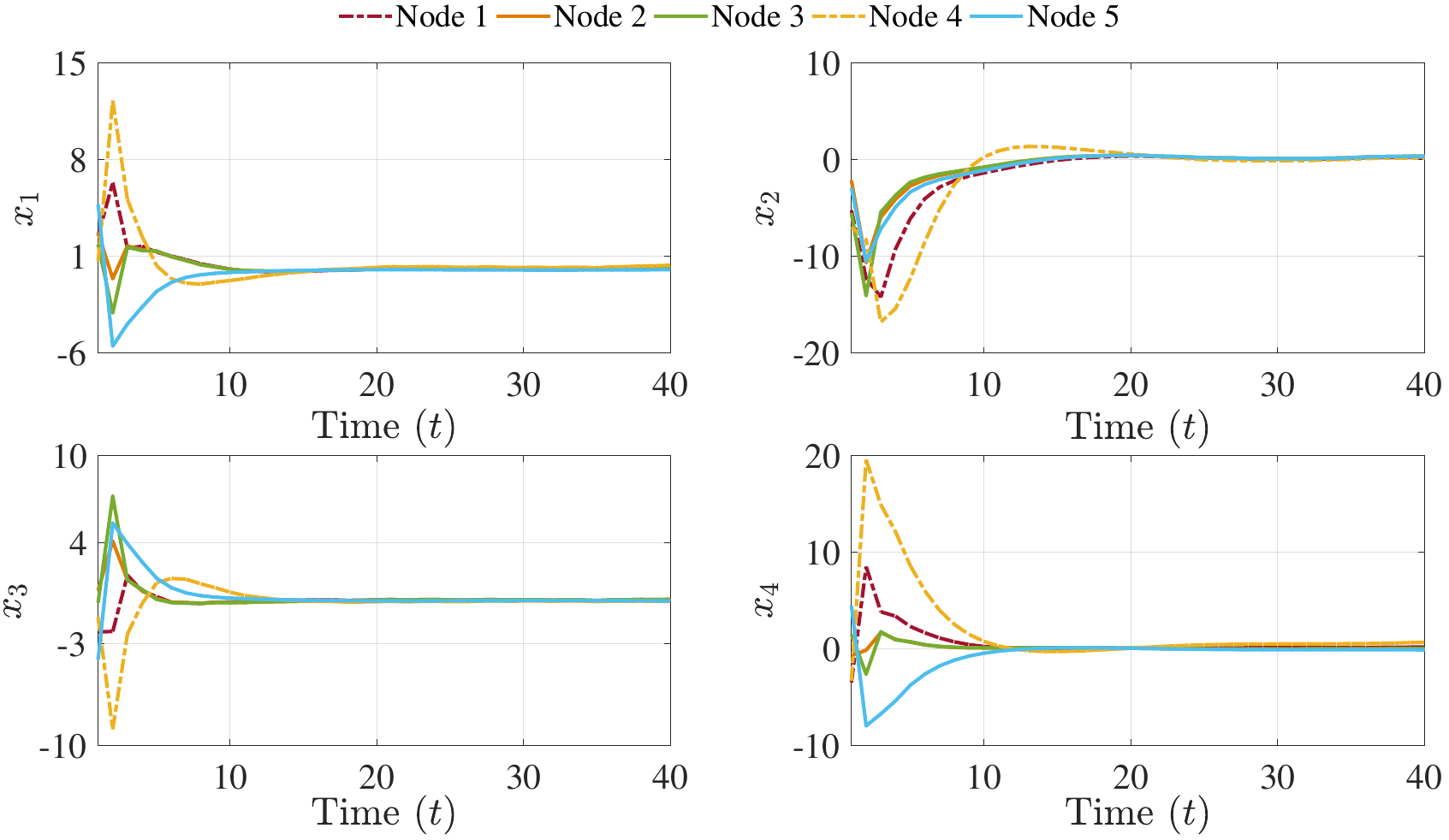}  
\caption{The estimation error of ID--DUIO.}
\label{fig.errID-DUIO}
\end{figure}

The model-based DUIO demonstrates superior estimation performance in Figs. \ref{fig.DUIO}--\ref{fig.errDUIO}. Figs. \ref{fig.ID-DUIO} --\ref{fig.errID-DUIO} illustrate that ID-DUIO exhibits inferior performance  compared to the other two methods. Due to the unknown disturbance in the offline data, it is  impossible to 
determine the original system model using identification methods. Therefore, the trajectories generated by the ID-DUIO are not fully compatible with the target system trajectories. The proposed D-DUIO method outperforms ID-DUIO and achieves competitive performance with the model-based DUIO.


To further compare the performance of the three methods, the evaluation metrics including mean-squared error (MSE) and mean-absolute error (MAE) are employed based on $100$ independent Monte Carlo experiments. 
The MSE of   sensor $i\in {\mathcal M}$  during   an experiment over the time interval $[1,T]$ is defined as ${\rm MSE}_i=(1/T){\int_0^{T}\left( x(\tau )-\hat{x}_i(\tau) \right)^2}d\tau $, where $x(\tau)$ is the true state of the target system, and $\hat x_i(\tau)$ is the state estimated by node $i$ during the experiment.
For convenience of notation, we denote by ${\rm MSE}_i^k$ the MSE of sensor $i$  during the $k$th experiment.
The MSE of all estimates during the $k$th experiment becomes ${\rm MSE}^k=(1/M)
\sum_{i=1}^{M}{\rm MSE}_i^k$. After $100$ independent experiments, the MSE becomes ${\rm MSE}=(1/100)
\sum_{k=1}^{100}{\rm MSE}^k$.  The MAE is defined analogously.

The proposed D-DUIO method shows a significant improvement in MSE and MAE relative to ID-DUIO. The difference between D-DUIO and DUIO in MSE and MAE of are $7.75\%$ and $1.85\%$, respectively. Compared to the other two DSE methods, the effectiveness of the proposed D-DUIO is demonstrated.
\begin{table}[ht]
\label{tab.one}	
\caption{Evaluation metrics of DUIO, D-DUIO and ID-DUIO}
\centering	
\begin{tabular}{cccc} 
\toprule
\multirow{1}{*}{Method}&{MSE}&{}&{MAE} \\
\midrule
\multirow{1}{*}{Model-based DUIO} & {5.1627}&{}&{0.5614}  \\ 

{Proposed D-DUIO}& \multirow{1}{*}{5.5629}&{}& \multirow{1}{*}{0.5718} \\

\multirow{1}{*}{ID-DUIO } &6.5461&{}&{0.8510}  \\
\bottomrule
\end{tabular}
\end{table}

\section{Conclusions}\label{sec.con}
In this paper, we investigated the problem of designing
a DUIO for a continuous-time  LTI system, subject to unknown inputs and
process disturbances, such that the state
estimation error asymptotically converges to zero.
First, we analyzed the problem using a model-based approach and derived a new sufficient condition to ensure that the state estimates of the proposed DUIO converge to the true system state asymptotically. Then, we proposed a novel D-DUIO to estimate the state of the unknown target system. We showed that, under mild assumptions, offline data are representative of any online input/output/state trajectory generated by the continuous-time unknown system.
In addition, it was shown that, using only the offline
data, it is possible to both  verify if the given sufficient condition for  the existence of a D-DUIO holds and to derive  a family of possible choices for  the D-DUIO matrices.
Simulation results validated the efficacy of the proposed approach. Future research will focus on extending the framework to more complex settings, such as nonlinear systems and switching network topologies.

\bibliographystyle{plainnat}
\bibliography{reference}

\clearpage

\renewcommand{\thesection}{Appendix A}
\section{Lemma \ref{small_lap} and its proof}\label{sec.App.Laplacian}
\begin{lemma}  \label{small_lap} 
	Let $i$ be arbitrary in ${\mathcal M}$, and let $\tilde {\mathcal L}$ be the $(M-1)\times (M-1)$ matrix obtained from ${\mathcal L}$ by removing the $i$th row and $i$th  column.
	Under Assumption \ref{Ass:network}, 
	$-\tilde{\mathcal L}$ is a Hurwitz compartmental matrix. 
\end{lemma}

\begin{pf}
	Suppose, without loss of generality, $i = 1$. We have already noticed (see Remark \ref{laplacian}) that $-{\mathcal L}$ is compartmental, irreducible and 
	${\mathbf 1}_M^\top(-\mathcal L) = {\bf 0}^\top$. Therefore, since $-\tilde{\mathcal L}$ is obtained from $-\mathcal L$ by removing its first row and  first column,
	$-\tilde{\mathcal L}$ is Metzler and satisfies
	${\mathbf 1}_{M-1}^\top(-\tilde{\mathcal L}) \le {\bf 0}^\top$.
	In addition, the irreducibility assumption on $\mathcal L$ implies that there exists $j\in\{2,\dots,M\}$ such that 
	$[-{\mathcal L}]_{1j} > 0$, and hence
	${\mathbf 1}_{M-1}^\top(-\tilde{\mathcal L}^\top)e_j < 0$.\\
	We now distinguish between two cases. \smallskip \\
	$\bullet \ \tilde{\mathcal L}$ {\em irreducible}. If $\tilde{\mathcal L}$ is irreducible, we can exploit Lemma 3 in \cite{SCL2018} to prove that $-\tilde{\mathcal L}$ is  Hurwitz. \smallskip \\
	$\bullet \ \tilde{\mathcal L}$ {\em reducible}. Let $P \in \mathbb R^{(M-1)\times (M-1)}$ be a permutation matrix such that 
	$$
	{P}^\top(-\tilde{\mathcal L}) P = \begin{bmatrix}
		Q_1 ~&~ ~&~ ~&~ \cr
		& Q_2 & & \cr
		& & \ddots & \cr 
		& & & Q_k
	\end{bmatrix},
	$$
	where $Q_i\in\mathbb R^{n_i\times n_i}$ is a Metzler irreducible matrix, and $\sum_{i=1}^{k}{n_i} = M-1$. 
	Then, 
	\begin{equation}\label{irred}
		{\!\!\!\!\left[\begin{array}{c|c}
				1 ~&~ \cr
				\hline
				~&~ {P}^\top
			\end{array}\right]\!\!(-\mathcal L)\!\! \left[\begin{array}{c|c}
				1~ &~ \cr
				\hline
				~&~ {P}
			\end{array}\right] \!=\! \!
			\left[\begin{array}{c|ccc}
				-l_{11} ~&~ \Delta_{1} ~& \dots &~ \Delta_{k} \cr
				\hline
				\Delta_{1}^\top & Q_1 & & \cr
				\vdots & & \ddots & \cr
				\Delta_{k}^\top & & & Q_k
			\end{array}\right]\!\!.}\!\!
	\end{equation}
	Since $-\mathcal L$ is irreducible, then the matrix in \eqref{irred} is irreducible too. Therefore, for every $j\in \{1,\dots,k\}$, $\Delta_{j} \ne 0$, and hence $\forall i \in \{1,\dots,k\}, \ {\mathbf 1}_{n_i}^\top Q_i \le {\bf 0}^\top$, with at least one entry strictly negative. Since $\forall i \in \{1,\dots,k\}$, $Q_i$ is irreducible, we can apply again Lemma 3 in \cite{SCL2018} to conclude that $Q_i$ is Hurwitz, and hence $-\tilde{\mathcal L}$ is Hurwitz. 
\end{pf}

\renewcommand{\thesection}{Appendix B}
\section{Derivation of Eqn.\eqref{eq:err5}}\label{sec.App.a}
\begin{pf}
Upon introducing the estimation
error $e_i(t)$ of node $i$, 
one deduces that the $i$th estimation error dynamics is
\begin{align}
e_i(t)&=x(t)-\hat x_i(t) \nonumber\\
&=x(t) - z_i(t) - H_iy_i(t) \nonumber\\
&= (I-H_iC_i)x_i(t)-z_i(t). \label{eq:errz1}
\end{align}
Taking the time derivative of \eqref{eq:errz1} yields
\begin{align}
&\dot e_i(t) =  (I-H_iC_i) \dot x_i(t)- \dot z_i(t)\nonumber\\
&= (I -  H_iC_i)(A x(t) + B_i^m u_i(t) + B_i^p w_i(t))- E_i z_i(t)  \nonumber\\
&\quad  -F_iu_i(t)-L_iy_i(t)-K_i\sum\limits_{j=1}^Ma_{ij}[\hat {x}_j(t)-\hat{x}_i(t)] \nonumber\\
&=  E_i e_i(t) + (I -  H_iC_i)(A x(t) + B_i^m u_i(t) + B_i^p w_i(t)) \nonumber\\
& \quad -F_iu_i(t)+E_iH_iy_i(t)-L_iy_i(t)-E_ix(t)\nonumber \\
& \quad  - K_i \sum\limits_{j=1}^Ma_{ij}[\hat{x}_j(t)-\hat{x}_i(t)]\nonumber \\
&= E_i e_i(t)+[(I -  H_iC_i)A-E_i(I-H_iC_i)-L_iC_i]x(t) \nonumber\\
& \quad +\  [(I - H_iC_i)B_i^m-F_i ]u_i(t) + (I -  H_iC_i)B_i^p w_i(t) \nonumber\\
& \quad  -K_i \sum\limits_{j=1}^Ma_{ij}[e_i(t)-e_j(t)]. \!\!\!\! \!\!\!\! \label{eq:errz}
\end{align}
So, by referring to $
e_G(t) =\big[  e^\top_1(t) \ \cdots \  e^\top_i(t) \ \cdots \   e^\top_M(t)  \big]^\top$, one can rewrite \eqref{eq:errz} compactly as \eqref{eq:err5}.
\end{pf}

\renewcommand{\thesection}{Appendix C}
\section{Proof of Theorem \ref{gamma_choice}}\label{sec.App.b}

\begin{pf} First of all, we observe that by choosing the matrices $E_i, F_i,$ and $L_i, i\in {\mathcal M}$, as in \eqref{eq:no1}-\eqref{eq:no5} we satisfy conditions \eqref{eq:equality}, and hence the estimation error evolves according to   equation
\eqref{eq:errvv}.
Moreover,  $E_1 =   (I-\bar H _1 C_1)A -   M_1 C_1$ is Hurwitz. Therefore, if we impose $K_1 = 0$ and $K_i = \gamma I, i = 2, \dots, M$, we obtain 
\begin{align*}
&{\rm diag}(E_i) - {\rm diag}(K_i) (\mathcal L \otimes I) = \\
& \begin{bmatrix}
E_1 &\vline& \cr
\hline
&\vline&\tilde E
\end{bmatrix} - 
\begin{bmatrix}
0 &\vline&  \cr
\hline
&\vline&  \gamma I_{(M-1)n_x}
\end{bmatrix}\!\!\!
\left[
\begin{array}{c|ccc}
l_{11}I ~&~  l_{12} ~& \dots &~ l_{1M} \cr 
\hline
l_{12}I & & &  \cr 
\vdots & \multicolumn{3}{c}{\smash{\raisebox{.5\normalbaselineskip}{$\tilde{\mathcal L} \otimes I$}}} \cr 
l_{1M}& & &
\end{array}
\right]  \!\!\! \\
&
=\left[
\begin{array}{c|c}
E_1 & 0 \cr 
\hline
* ~&~\tilde E - \gamma (\tilde{\mathcal L} \otimes I) 
\end{array}\right].
\end{align*}
Now it remains to prove that we can always choose $\gamma >0$ so that $\tilde E - \gamma (\tilde{\mathcal L} \otimes I)$ is Hurwitz. Consider the following Lyapunov function 
$V(\tilde e(t)) = \tilde e^\top(t) \tilde e(t),
$
which is a positive definite function of $\tilde e(t) \triangleq \begin{bmatrix}
e_2^\top(t) & \dots & e_M^\top(t)
\end{bmatrix}^\top$, whose dynamics is given by 
\begin{equation} \label{small_dyn}
\dot{\tilde e}(t) = (\tilde E - \gamma (\tilde{\mathcal L} \otimes I)) \tilde e(t).
\end{equation}
The time derivative of $V$ along \eqref{small_dyn} satisfies 
\begin{align*}
\dot V(\tilde e(t))&= \tilde e^\top(t) (\tilde E + \tilde E^\top) \tilde e(t) - 2 \gamma \tilde e^\top(t)(\tilde{\mathcal L} \otimes I) \tilde e(t) \\
&\le (\Vert \tilde E + \tilde E^\top \Vert - 2\gamma \lambda_{min}(\tilde{\mathcal L} \otimes I)) \Vert \tilde e(t) \Vert^2,
\end{align*}
and hence if $\gamma$ satisfies   \eqref{eq:gammamin}
then $\tilde E - \gamma (\tilde{\mathcal L} \otimes I)$ is Hurwitz. Consequently, 
${\rm diag}(E_i) - {\rm diag}(K_i) (\mathcal L \otimes I) $ is Hurwitz, and the state estimation error asymptotically converges to zero.  
\end{pf}

\renewcommand{\thesection}{Appendix D}
\section{Proof of Theorem \ref{The.1}}\label{sec.App.c}
\begin{pf}
Consider any input/output/state trajectory of ${\mathcal T}_i$,   described in vector form as $[
u^\top_i(t) \ y^\top_i(t) \ \dot {y}^\top_i(t)$  $x^\top(t)  \ \dot  x^\top(t) ]^\top$.
From \eqref{eq:systemi}--\eqref{eq:output} one deduces that 
\begin{equation}
\begin{bmatrix}
u_i(t)  \\ y_i(t) \\ \dot {y}_i(t)  \\  x(t)  \\ \dot x(t) \\
\end{bmatrix}
= \Theta_i   \begin{bmatrix}
u_i(t)  \\ w_i(t)  \\	x(t)  \\
\end{bmatrix}, \label{eq:substitu}
\end{equation}
where 
$$ \Theta_i :=  \begin{bmatrix}
I & \mathbf 0 & \mathbf 0 \\
\mathbf	0 &  \mathbf	0 & C_i  \\
C_iB_i^m     ~&~ C_iB_i^p      ~&~ C_i A  \\
\mathbf	0  & \mathbf	0  &  I  \\
B_i^m  & B_i^p & A \\
\end{bmatrix}.$$
Therefore an input/output/state trajectory belongs to $T_{{\mathcal T}_i}$ 
if and only if it can be expressed as in \eqref{eq:substitu} for every $t\in {\mathbb R}_+$.
Indeed, 
given any sequence $\{[u_i^\top(t) \ w_i^\top(t) ]^\top\}_{{ t\in {\mathbb R}_+}}$ and any initial state $x(0)$, the sequence $(\{u_i(t) \}_{{ t\in {\mathbb R}_+}},
\{ y_i(t)\}_{{ t\in {\mathbb R}_+}},$
$\{ \dot {y}_i(t)\}_{{ t\in {\mathbb R}_+}},$
$\{ x(t)\}_{{ t\in {\mathbb R}_+}}, \{\dot x(t)\}_{{ t\in {\mathbb R}_+}})$ obtained by iteratively solving 
\eqref{eq:substitu} for $t \in {\mathbb R}_+$, is  a trajectory of ${\mathcal T}_i$.
Conversely, every sequence 
$(\{u_i(t) \}_{{ t\in {\mathbb R}_+}},\{ y_i(t)\}_{{t\in {\mathbb R}_+}}, \{ \dot {y}_i(t)\}_{{ t\in {\mathbb R}_+}},$
$\{ x(t)\}_{{ t\in {\mathbb R}_+}}, \{\dot x(t)\}_{{ t\in {\mathbb R}_+}})$ obtained by iteratively solving 
\eqref{eq:substitu} for $t \in {\mathbb Z}_+$ is  a trajectory of ${\mathcal T}_i$, corresponding to  some sequence {$\{[u_i^\top(t) \ w_i^\top(t) ]^\top\}_{{ t\in {\mathbb R}_+}}$} and some initial state $x(0)$.
On the other hand, the  historical data $(\bar{u}_i,\bar{y}_i,\dot {\bar{y}}_i, \bar{x}_i, \dot{\bar{x}}_i)$ collected by node $i$ are generated by ${\mathcal T}_i$ 
and hence satisfy 
\begin{equation}
	\begin{bmatrix}
		U_i  \\ Y_i  \\ {\dot{Y}_i}  \\ X_i  \\  {\dot{X}_i}  \\
	\end{bmatrix} = {\Theta_i} \!   \begin{bmatrix}
		U_i  \\ W_i  \\ X_i  \\
	\end{bmatrix}.
	\label{eq:hisPR}
\end{equation} 
We are now in a position to prove the theorem statement. As the system is linear, for every 
vector $ g_i(t)\in {\mathbb R}^N$ we have  
$$\begin{bmatrix}
	u_i(t)  \\ y_i(t)  \\ 
	\dot y_i(t)  \\ x(t)  \\  \dot x(t)  \\
\end{bmatrix}   = \begin{bmatrix}
	U_i  \\ Y_i   \\ {\dot{Y}_i}  \\ X_i \\ {\dot{X}_i}  \\
\end{bmatrix}   g_i(t)   = {\Theta_i} \!   \begin{bmatrix}
	U_i  \\ W_i  \\ X_i  \\ 
\end{bmatrix}   g_i(t) =   {\Theta_i} \! \begin{bmatrix}
	u_i(t)  \\ w_i(t)  \\ x(t)  \\ 
\end{bmatrix}.
$$
This proves that $T(\bar u_i,\bar y_i, \dot{\bar y}_i,\bar x_i,\dot {\bar x}_i)\subseteq T_{{\mathcal T}_i}$. To prove that  $T_{{\mathcal T}_i}\subseteq T(\bar u_i,\bar y_i, \dot{\bar y}_i,\bar x_i,\dot {\bar x}_i)$, we first recall from Assumption \ref{Ass:hisiuxy} that the matrix 
$ \begin{bmatrix}
	U_i^\top  & W_i^\top  &  X_i^\top  
\end{bmatrix}^\top$
has full row rank. 
Therefore, { for every $t\in {\mathbb R}_+$}, there exists a   vector $ g_i(t) \in {\mathbb R}^N$ such that
\begin{align*}
	\begin{bmatrix}
		u_i(t)  \\ w_i(t)  \\ x(t)  \\ 
	\end{bmatrix}  
	&= \begin{bmatrix}
		U_i  \\ W_i  \\ X_i  \\
	\end{bmatrix}  g_i(t).
\end{align*}
On the other hand, being a trajectory of  $T_{{\mathcal T}_i}$, it also satisfies
\eqref{eq:substitu}. By making use of \eqref{eq:hisPR}, then, we deduce that  ${ \forall t\in {\mathbb R}_+}$
\begin{equation}
	\begin{bmatrix}
		u_i(t)  \\ y_i(t)   \\ 
		\dot y_i(t) \\ x(t)  \\ \dot x(t)  \\
	\end{bmatrix}   =  \begin{bmatrix}
		U_i  \\ Y_i   \\ {\dot{Y}_i} \\ X_i  \\ {\dot{X}_i}  \\
	\end{bmatrix}    g_i(t),
	\label{eq:guyx}
\end{equation}
and hence
$T_{{\mathcal T}_i}\subseteq T(\bar u_i,\bar y_i, \dot {\bar y}_i,\bar x_i,\dot {\bar x}_i)$, which completes the proof.
\end{pf}

\renewcommand{\thesection}{Appendix E}
\section{Proof of Lemma \ref{Lemma.DD}}\label{sec.App.d}

\begin{pf} i)\ By making use of \eqref{eq:hisPR}, we deduce that 
$$\begin{bmatrix} U_i  \\  \dot{Y}_i  \\ X_i  
\end{bmatrix} =   \begin{bmatrix}
	I & \mathbf 0 & \mathbf 0 \\
	C_iB_i^m     & C_iB_i^p      & C_i A  \\
	\mathbf	0  & \mathbf	0  &  I  	\end{bmatrix} 
\begin{bmatrix}
	U_i  \\ W_i  \\ X_i  \\
\end{bmatrix},$$
as well as
$$ \begin{bmatrix}U_i  \\     X_i  \\ \dot{X}_i  
\end{bmatrix} = \begin{bmatrix}
	I & \mathbf 0 & \mathbf 0 \\
	\mathbf	0  & \mathbf	0  &  I  \\
	B_i^m  & B_i^p & A \\
\end{bmatrix}     \begin{bmatrix}
	U_i  \\ W_i  \\ X_i  \\
\end{bmatrix}.$$ 
By leveraging   Assumption
\ref{Ass:hisiuxy}  we can claim that 
$${\rm rank} \left(\begin{bmatrix} U_i  \\  \dot{Y}_i  \\ X_i  
\end{bmatrix} \right) = n_{m_i}+n_x + {\rm rank} (C_i B_i^p),$$
while 
$${\rm rank} \left(\begin{bmatrix} U_i  \\   X_i  \\\dot{X}_i  \\
\end{bmatrix} \right) = n_{m_i}+n_x + {\rm rank} (B_i^p).$$ 
Therefore, 
${\rm rank} (C_i B_i^p)= {\rm rank}(B_i^p)$   if and only if
\eqref{eq:2ranks} holds.

\noindent ii) \ The proof is inspired by the one for reconstructibility first provided in
Proposition   7 of \cite{FDI_Giulio_arXiv} 
and by the proofs of Theorems 1 and 2 in \cite{Darouach}, and hence it is concise.
We preliminarily observe that ${\rm rank} (\bar H_i) = {\rm rank} (B_i^p(C_iB_i^p)^\dagger)) = r_i$ and hence (see Lemma 11 in \cite{FDI_Giulio_arXiv}) 
${\rm rank} (I-\bar H_iC_i) = n_x- r_i$, as well as ${\rm ker} (I-\bar H_iC_i) = {\rm range} (B_i^p)$. This implies that
\begin{equation}
	\label{eq:tech1}
	\begin{bmatrix} I-\bar H_iC_i \cr (B_i^p)^\dagger\end{bmatrix}
\end{equation} 
is of full column rank.
By making use   of \eqref{eq:hisPR}, we can claim that
$$\begin{bmatrix}
	sX_i- \dot{X}_i \\
	U_i \\
	Y_i
\end{bmatrix} = 
\begin{bmatrix} 
	- B_i^m & - B_i^p  & sI -A \cr
	I_{m_i} & {\bf 0} & {\bf 0}\cr
	{\bf 0}& {\bf 0} & C_i
\end{bmatrix} \begin{bmatrix}
	U_i \\
	W_i\\
	X_i
\end{bmatrix}, $$
and hence, by Assumption \ref{Ass:hisiuxy}, condition \eqref{eq:DDdetecti} holds if and only if for every $s\in {\mathbb C}, {\rm Re}(s) \ge 0$,
\begin{equation}
	\label{eq:tech2}
	{\rm rank}\left(\begin{bmatrix}
		sI -  A & - B_i^p\\
		C_i & {\bf 0}
	\end{bmatrix}\right) =  n_x + r_i.
\end{equation}
By premultiplying 
the matrix in \eqref{eq:tech2} by the (full column rank) matrix 
$$
\begin{bmatrix} I-\bar H_iC_i  & {\bf 0} \cr (B_i^p)^\dagger & {\bf 0} \cr
	{\bf 0} & I\end{bmatrix},
$$
we deduce  (see the proof of Theorem 2 in \cite{Darouach}) that
condition \eqref{eq:tech2} holds if and only if  
$${\rm rank}\left(\begin{bmatrix}
	s (I-\bar H_i C_i) - (I-\bar H_i C_i)A\\
	C_i
\end{bmatrix}\right) =  n_x.$$
On the other hand, see the proof of Theorem 1 in \cite{Darouach},  the previous rank condition holds if and only if 
$${\rm rank}\left(\begin{bmatrix}
	s I- (I-\bar H_i C_i)A\\
	C_i
\end{bmatrix}\right) =  n_x.$$
Finally, by resorting to the PBH observability test, we can claim that 
$${\rm rank}\left(\begin{bmatrix}
	sI - (I-\bar H_i C_i)A\\
	C_i
\end{bmatrix}\right) =  n_x, \quad \forall s\in \mathbb C, {\rm Re}(s) \ge 0,$$
if and only if the pair $((I - \bar H_i C_i)A, C_i)$ is detectable.
So, to conclude, Assumption \ref{Ass:detect} holds  if and only if there exists $i\in {\mathcal M}$ such that
\eqref{eq:DDdetecti} holds.
\end{pf}

\renewcommand{\thesection}{Appendix F}
\section{Proof of Theorem \ref{The.2}}\label{sec.App.e}
\begin{pf}
i)\  We first notice that since data have been collected from the real system, then
they satisfy equation \eqref{eq:systemi}. Therefore for every $i\in {\mathcal M}$ we have
\begin{equation}
	\label{eq:DDsys}
	\dot{X}_i = A X_i + B_i^m U_i + B_i^p W_i.
\end{equation}
As  condition 
\eqref{eq:2ranks} holds for every $i\in {\mathcal M}$, this means that ${\rm rank}(C_i B_i^p)= {\rm rank} (B_i^p)=r_i$, or equivalently 
(see Lemma \ref{lem.con}) that there exists $H_i$ such that $H_i C_i B_i^p = B_i^p$.
We assume that $H_i=\bar H_i = B_i^p(C_i B_i^p)^\dagger$ and hence it has rank $r_i$.
If  we premultiply \eqref{eq:DDsys} by $(I-\bar H_i C_i)$ we get
$$(I-\bar H_i C_i)\dot{X}_i = (I-\bar H_i C_i)A X_i + (I-\bar H_i C_i)B_i^m U_i,$$
from which we deduce that
$$\dot{X}_i = \begin{bmatrix}
	(I-\bar H_i C_i)B_i^m& \bar H_i &(I-\bar H_i C_i)A\end{bmatrix} \begin{bmatrix} U_i\cr \dot{Y}_i \cr X_i\end{bmatrix}.$$
So, \eqref{eq:mainrel} holds for every $i\in {\mathcal M}$, with
$$T_u^i := (I-\bar H_i C_i)B_i^m, \quad
T_y^i := \bar H_i, \quad T_x^i := (I-\bar H_i C_i)A.$$
On the other hand if condition 
\eqref{eq:DDdetecti} holds for $i=1$, then $(T_x^1,C_1) = ((I-\bar H_1 C_1)A, C_1)$ is detectable, by Lemma \ref{Lemma.DD}.
Therefore i) holds true.

\noindent ii)\ Follows from Lemma 9 in \cite{FDI_Giulio_arXiv}. 

\noindent iii)\ Consider now any family of solutions
$T_u^i, T_y^i,$ and $T_x^i$,  with 
${\rm rank} (T_y^i)= r_i$,  $i\in {\mathcal M}$, of equations
\eqref{eq:mainrel}.
Assume that the matrices $E_i, F_i, H_i, L_i$ and $K_i, i\in {\mathcal M}$, are defined as in \eqref{eq:conditionsDD}, and set  
$E_G := {\rm diag}(E_i)$, $F_G := {\rm diag}(F_i)$, $L_G := {\rm diag}(L_i)$, $H_G := {\rm diag}(H_i)$ and $K_G := {\rm diag}(K_i)$; $z_G(t) := \big[ 
z^\top_1(t) \ \cdots \ z^\top_i(t) \ \cdots \  z^\top_M(t)  \big]^\top$. The vectors $u_G(t),~y_G(t),~\hat x_G(t),~\dot z_G(t)$ are defined in an analogous way. 
We want to prove that
\begin{equation}
	{\text{DUIO$_G$}}\!
	\left\{\begin{array}{l}
		\begin{aligned}
			\!\! \dot z_G(t)
			&=E_G z_G(t)+F_G u_G(t)+L_G y_G(t)\\
			&\quad -K_G (\mathcal L \otimes I) \hat x_G(t)\\
			\!\!	\hat x_G(t)&=z_G(t)+H_G y_G(t),
		\end{aligned}
	\end{array}\right.
	\label{eq:duioG}
\end{equation}
and that $E_G  -K_G (\mathcal L \otimes I)$ is Hurwitz.

For $i=2,\dots,M$, equation \eqref{eq:mainrel}  can be rewritten as
$$\dot{X} _i = \begin{bmatrix} F_i &   H_i & E_i\end{bmatrix} \begin{bmatrix}U_i\cr \dot{Y}_i\cr X_i\end{bmatrix}
= \begin{bmatrix} F_i &   0 & H_i & E_i\end{bmatrix} \begin{bmatrix}U_i\cr Y_i\cr \dot{Y}_i\cr X_i\end{bmatrix},$$
while for $i=1$ we have 
\begin{eqnarray*}
	\dot{X} _1 &=& \begin{bmatrix} F_1 &    M_1 & H_1 & E_1\end{bmatrix} \begin{bmatrix}U_1\cr Y_1\cr \dot{Y}_1\cr X_1\end{bmatrix}
	\\
	&=& \begin{bmatrix} F_1 &    L_1 - E_1 H_1 & H_1 & E_1\end{bmatrix} \begin{bmatrix}U_1\cr Y_1\cr \dot{Y}_1\cr X_1\end{bmatrix}.
\end{eqnarray*}

By Theorem \ref{The.1}, we know that for every $i\in {\mathcal M}$ and every $t\in {\mathbb R}_+$, 
$\begin{bmatrix}
	u_i^\top(t)  \ y_i^\top(t)  \ x^\top(t)  \ 
	\dot y_i^\top(t)  \ \dot x^\top(t) 
\end{bmatrix}^\top  = \begin{bmatrix}
	U_i^\top  \ Y_i^\top  \ X_i^\top  \ {\dot{Y}_i}^\top  \ {\dot{X}_i}^\top
\end{bmatrix}^\top \!\! g_i(t),$ for some vector $g_i(t)$. 

So, upon setting \begin{align*}
		x_G(t)&:=\left[ \begin{matrix}
			x^\top(t) \ \cdots \ x^\top(t) \ \cdots \ x^\top(t)  \\
		\end{matrix} \right]^\top = {\mathbf 1}_M \otimes x(t), \\
		{u}_G(t)&:=  \begin{bmatrix}
			u_1^\top(t) \ \cdots \ u_i^\top(t) \ \cdots  \ u_M^\top(t)  \\
		\end{bmatrix}^\top,\\
		{y}_G(t)&:=  \begin{bmatrix}
			y_1^\top(t) \ \cdots \ y_i^\top(t) \ \cdots  \ y_M^\top(t)  \\
		\end{bmatrix}^\top,
	\end{align*}
	and, by analogously defining the vectors $\dot x_G(t)$ and $\dot y_G(t)$,
	the previous identities lead to 
	\begin{align}
		\dot x_G(t) &= E_G x_G(t) + F_G u_G(t) 
		+(L_G-E_G H_G) y_G(t)	\nonumber\\
		&  +H_G \dot y_G(t).  \label{eq:xk1nocons}
	\end{align} 
	We observe that since all blocks in $x_G(t)$ are identical, 
	$x_G(t)$  belongs to the kernel of the Laplacian $\mathcal L$,
	therefore  we can   substitute \eqref{eq:xk1nocons} with the following equation 
	\begin{align}
		\dot x_G(t) &= F_G u_G(t)
		+H_G \dot y_G(t)
		+(L_G-E_G H_G)y_G(t) 	\nonumber\\
		& +(E_G - K_G(\mathcal{L} \otimes I))x_G(t). \label{eq:xk1}
	\end{align}
	Upon defining
	\begin{align}
		\dot z_G(t)
		& :=F_G u_G(t)
		+(L_G-E_G H_G) y_G(t)	\nonumber\\
		&+(E_G - K_G(\mathcal{L} \otimes I))x_G(t),
		\label{eq:zk1}
	\end{align}
	we have that \eqref{eq:duioG} holds.
	Finally, the fact that $E_G -K_G(\mathcal L \otimes I)$ is Hurwitz stable follows    from Theorem \ref{gamma_choice}.
\end{pf}	
\end{document}